\documentclass[journal]{IEEEtran}
\ifCLASSINFOpdf
\else
\fi
\usepackage{graphicx}
\usepackage{setspace}
\usepackage{amsmath}
\usepackage{amssymb} 
\usepackage{bm} 
\usepackage{cite}
\usepackage{float}
\usepackage[usenames,dvipsnames]{pstricks}
\usepackage{epsfig}
\usepackage{pst-grad} % For gradients
\usepackage{pst-plot} % For axes
\usepackage{tabularx}
\usepackage{longtable}
\usepackage{multicol}
\usepackage{supertabular}
\usepackage{verbatim}
\usepackage{lipsum}  % for filler text
\usepackage[keeplastbox]{flushend}
\usepackage{booktabs}
\usepackage{ltxtable}
\usepackage{afterpage}
\newcommand{\lb} {\left}
\newcommand{\rb} {\right}
\newcommand{\nn} {\nonumber}

\allowdisplaybreaks

\begin{document}

% \title{Secrecy Analysis of Threshold-Selection DF Relaying With and Without CSI Knowledge}
\title{Effects of CSI Knowledge on Secrecy of Threshold-Selection Decode-and-Forward Relaying}

\author{Chinmoy Kundu,~\IEEEmembership{Member,~IEEE,}
        Sarbani Ghose,~\IEEEmembership{Member,~IEEE,}       
        Telex M. N. Ngatched,~\IEEEmembership{Member, IEEE,} 
        Octavia A. Dobre ~\IEEEmembership{Senior Member, IEEE},
        Trung Q. Duong ~\IEEEmembership{Senior Member, IEEE},
         and
        Ranjan Bose,~\IEEEmembership{Senior Member, IEEE }
        %         and~Jane~Doe,~\IEEEmembership{Life~Fellow,~IEEE}% <-this % stops a space

        \thanks{This work was presented in part at the 
          2016 IEEE 84th Vehicular Technology Conference, Montreal, Canada.}

        \thanks{Chinmoy Kundu and Trung Q. Duong are with the School of Electronics, Electrical Engineering and Computer Science, 
Queen's University Belfast, Northern Ireland, U.K.,  e-mail: \{c.kundu, trung.q.duong\}@qub.ac.uk.}
\thanks{Sarbani Ghose is with the Cryptology
and Security Research Unit, Indian Statistical Institute, Kolkata 700108, India, 
e-mail: sarbani\_v@isical.ac.in.}
\thanks{Telex M. N. Ngatched and Octavia A. Dobre are with the Engineering and Applied Science, Memorial University, Canada,
e-mail: tngatched@grenfell.mun.ca, odobre@mun.ca.}% <-this % stops a space
\thanks{Ranjan Bose is with the Department 
of Electrical Engineering, Indian Institute of Technology Delhi, New Delhi-110016, India, 
e-mail: rbose@ee.iitd.ac.in.}

\thanks{This work was supported in part by the Royal Society-SERB Newton International Fellowship under Grant NF151345, 
the U.K. Engineering and Physical Sciences Research Council under Grant EP/P019374/1,
the DST-SERB under National Post Doctoral Fellowship Grant PDF/2016/003637, and the 
Natural Science and Engineering Council of Canada (NSERC), through its Discovery program.} 
}% <-this % stops a space
% \thanks{Manuscript received April 19, 2005; revised December 27, 2012.}}
\maketitle

\begin{abstract}
This paper considers secrecy of a three node cooperative wireless system in the presence of
a passive eavesdropper. The threshold-selection decode-and-forward (DF) relay is considered, 
which can decode the source message correctly only if a predefined signal-to-noise ratio (SNR) 
is achieved. The effects of channel state information (CSI) availability on secrecy outage probability (SOP) 
and ergodic secrecy rate (ESR) are investigated, and closed-form expressions are derived.
Diversity is achieved from  the direct and relaying paths both at the destination and at the 
eavesdropper by combinations of maximal-ratio combining (MRC) and selection combining (SC) schemes. 
An asymptotic analysis is provided when each hop SNR is the same in the balanced case and when it is 
different in the unbalanced case. The analysis shows that both hops can be a bottleneck 
for secure communication; however, they do not affect the secrecy identically. 
While it is observed that CSI knowledge can improve secrecy, the amount of 
improvement for SOP is more when the required rate is low and for ESR when the operating SNR is also low.
It is also shown that the source to eavesdropper link SNR is more crucial for secure communication.
\end{abstract}

\begin{IEEEkeywords} 
Channel state information, cooperative diversity, decode-and-forward relay, ergodic secrecy rate, 
secrecy outage probability, threshold-selection relay.  
\end{IEEEkeywords}
% % \IEEEpeerreviewmaketitle

\section{Introduction}
\label{sec_intro}

Due to the inherent openness and broadcast nature of the transmission medium, wireless
communications systems are particularly vulnerable to eavesdropping. Any unintended receiver
within the range of a transmitting antenna can overhear and decode the transmitted signal,
compromising the system security \cite{wyner_wiretap, Csiszar_broad_ch_with_conf, hellman_gaussian_wiretap}. 
Traditionally, security issues have been
dealt with at upper-layers of the communication protocol stack using cryptographic techniques.
Although cryptographic methods have proven to be efficient, they rely on the assumed limited
computing capabilities of the eavesdroppers and exhibit vulnerabilities in terms of the inevitable
secret key distribution as well as management. Introduced by Wyner, physical layer security
(PLS) has emerged as a promising technique to complement cryptographic methods, and significantly 
improve the security of wireless networks \cite{Rodrigues_sec_cap_wire_ch, McLaughlin_wireless_info_theo_sec, 
Gamal_On_the_Sec_Cap_Fad_Ch, poor_infor_theo_sec}. Unlike cryptographic approaches, PLS
exploits the physical layer properties of the communication system to maximize the uncertainty
concerning the source message at the eavesdropper.

When the source-destination channel is weaker than the source-eavesdropper channel,
positive secrecy rate can be achieved using a multiple transmit antenna system by improving 
the diversity gain of the legitimate link. An alternative solution to avoid the use of 
complex multiple antenna system is
to use cooperative relaying techniques \cite{Laneman_Wornell_cooperative_diversity}, 
as initially proposed by the authors in 
\cite{Gamal_The_Relay_Eaves_Ch_Coop_Sec}. Since then,
various cooperative relaying strategies, namely, amplify-and-forward (AF), decode-
and-forward (DF), noise forwarding, compress-and-forward, along with jamming techniques have
been investigated for secrecy enhancement \cite{Petropulu_Poor_Impr_Wire_Phylay_Sec, 
Petropulu_On_Coop_Rel_Scheme, Liu_rel_place_PLS}. However, thanks to their ability to resist noise
propagation to subsequent stages, DF relays have gained more importance in PLS.

Early works on cooperative techniques to improve the secrecy performance of wireless systems 
\cite{krikidis_twc_Rel_Sel_Jam, 
krikidis_iet_opport_rel_sel, 
Zou_Wang_Shen_optimal_relay_sel,
Bao_Relay_Selection_Schemes_Dual_Hop_Security, 
Alotaibi_Relay_Selection_MultiDestination, 
Poor_Security_Enhancement_Cooperative,
Hui_Secure_Relay_Jammer_Selection,
Kundu_relsel} assumed that the source had no direct link with the destination and the eavesdropper,
thereby indicating that the direct links were in deep shadowing. This assumption was slightly
relaxed in \cite{Nosrati_Wang_Secrecy_capacity_enhancement_two_hop,
Aissa_CCI_relay_selection,
sarbani_Kundu_threshold_relay, 
Cai_Cai_Average_secrecy_rate_analysis},
where only the direct link from source to destination was neglected.
The more practical scenario, which includes the direct links from the source to destination and 
eavesdropper, was recently considered in 
\cite{Alotaibi_Ergodic_Secrecy_Capacity_Selection,
Lei_fan_Secure_multiuser_communications,
Yeoh_Yang_SOP_of_Selective_Relaying,
Qahtani_relsel, 
kundu_globecom16,
kundu_vtcfall_16}. 
In the presence of direct links, both the
destination and the eavesdropper have access to two independent versions of the source message
and can therefore apply diversity combining techniques. Direct and relayed links are combined
at the eavesdropper using maximal ratio combining (MRC) and selection combining (SC) in 
\cite{Aissa_CCI_relay_selection}, and MRC in 
\cite{Nosrati_Wang_Secrecy_capacity_enhancement_two_hop, 
 sarbani_Kundu_threshold_relay, Cai_Cai_Average_secrecy_rate_analysis}. 
 Diversity combining is performed both at the destination
and eavesdropper using MRC technique in 
\cite{Alotaibi_Ergodic_Secrecy_Capacity_Selection, 
Lei_fan_Secure_multiuser_communications, 
kundu_globecom16, 
kundu_vtcfall_16}. Diversity is obtained by SC
at the destination with MRC at the eavesdropper in \cite{Yeoh_Yang_SOP_of_Selective_Relaying}. 
In \cite{Qahtani_relsel}, MRC, distributed selection
combining (DSC), and distributed switched and stay combining (DSSC) schemes are considered
at the destination along with MRC at the eavesdropper.

On the other hand, initial works on PLS in DF relay cooperative systems only considered the
high signal-to-noise ratio (SNR) regime for the source to relay link 
\cite{krikidis_iet_opport_rel_sel, 
krikidis_twc_Rel_Sel_Jam, 
Bao_Relay_Selection_Schemes_Dual_Hop_Security, 
Alotaibi_Relay_Selection_MultiDestination, 
Poor_Security_Enhancement_Cooperative, 
Hui_Secure_Relay_Jammer_Selection}. 
Though this assumption simplifies the analysis, it is not very practical as fading can severely degrade
the channel quality of a link in wireless communication systems. Such degradation can induce
decoding errors at the relay, leading to a significant reduction of the SNR at the destination
if diversity combining is performed. In
\cite{Nosrati_Wang_Secrecy_capacity_enhancement_two_hop,
Zou_Wang_Shen_optimal_relay_sel, 
Aissa_CCI_relay_selection, 
Kundu_relsel, 
Yeoh_Yang_SOP_of_Selective_Relaying,
Cai_Cai_Average_secrecy_rate_analysis}, the source to relay channel
quality is included in the secrecy analysis by assuming that the source-relay-destination branch
SNR is affected by the lowest quality hop of that particular branch, i.e., the minimum of the
source to relay and relay to destination link SNR. To better address the impact of the source to
relay link on the secrecy analysis, threshold-selection DF relay \cite{liu_opt_threshold}, 
in which perfect decoding is only possible if the instantaneous SNR exceeds a threshold, 
was recently introduced in \cite{sarbani_Kundu_threshold_relay, kundu_vtcfall_16, kundu_globecom16}.
In addition to this, still, effects of channel state information (CSI) knowledge at the transmitters 
on the secrecy of the relayed communication systems is not studied extensively. 
If CSI is available at the source, positive secrecy can be achieved even if the eavesdropper's 
link quality is better than the main link quality. However, when the CSI is not available
at the source, and instead, available only at the receiver, positive secrecy may not be guaranteed 
\cite{Olabiyi_Sec}. In \cite{Olabiyi_Sec}, only ergodic secrecy rate (ESR) is evaluate in 
the wiretap channel model. Recently, 
our works in \cite{sarbani_Kundu_threshold_relay, kundu_vtcfall_16} 
studied effect of CSI knowledge on both the secrecy outage probability (SOP) 
and ESR in communications using relay. 
In \cite{sarbani_Kundu_threshold_relay}, direct link was not considered 
from source to destination and \cite{kundu_vtcfall_16} only considered a single diversity scheme.

In this paper, we propose a detailed and comprehensive secrecy analysis of a single relay
system consisting of a source, a DF relay, a destination, and a passive eavesdropper. To account
for the first hop link quality and the effects of possible decoding errors on diversity combining,
threshold-selection DF relay is considered. From the proposed generalized system model, 
the particular cases of perfect decoding and basic wiretap channel can be obtained by setting
the threshold at the relay to zero and infinity, respectively. The joint impact of the direct and
relay links is taken into account and two important diversity techniques, namely, MRC and
SC, are considered with all possible combinations at both destination and eavesdropper
simultaneously. The effects of CSI knowledge at the transmitting nodes on the
SOP and ESR are thoroughly investigated
and closed-form expressions are derived in each case. Considering the cases when both
hops have the same average SNR (balanced case) and different average SNR (unbalanced case),
an asymptotic analysis is provided. It is shown that though both hops constitute a bottleneck for
secrecy, their effects are not identical.

The remainder of the paper is organized as follows. Section \ref{sec_system} describes the system model. The 
closed-form expressions of the SOP and ESR for the various diversity combination schemes performed
at the destination and eavesdropper are derived in Sections \ref{sec_sop} and \ref{sec_erg_sec_rate}, 
respectively. Section \ref{sec_asymp_ana} examines the asymptotic analysis 
of the SOP and ESR, while Section \ref{sec_results} presents numerical results. 
Finally, conclusions are provided in Section \ref{sec_conclusion}.

\textit{Notation:} $\mathbb{P}[\cdot]$ is the probability of occurrence of an event. 
For a random variable $X$,  $\mathbb{E}_X[\cdot]$ denotes expectation or mean of 
$X$, $F_X (\cdot)$ denotes its cumulative distribution function (CDF) and 
$f_X (\cdot)$ denotes the corresponding probability density function (PDF). 
$(x)^+\triangleq \max(0,x)$, and $\max{(\cdot)}$ and $\min{(\cdot)}$ denote the maximum 
and minimum of their arguments, respectively. 

\section{System Model}
\label{sec_system}
\begin{figure} 
\centering
\includegraphics[width=0.275\textwidth] {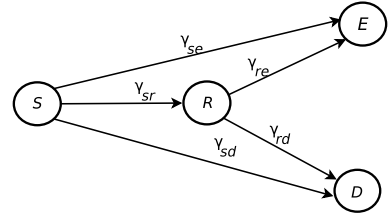}
\vspace*{-0.3cm}
\caption{System model for the threshold-selection relaying.}
\vspace*{.3cm}
\label{FIG_1}
\vspace{-1cm}
\end{figure}

The system model consists of a cooperative wireless network with a source ($S$), 
a relay ($R$) and a destination ($D$), along with a passive eavesdropper ($E$), all having 
single antenna, as shown in Fig. \ref{FIG_1}. $S$ broadcasts its message in the first time 
slot which is received by $R$, $D$, and $E$. If $R$ is able to decode the message correctly, 
then it would retransmit it in the second time slot.  $R$ can correctly decode the 
message only if a certain SNR threshold, $\gamma_{th}$, is satisfied. 
We assume that $S$ and $R$ use the same codebook for encoding the message. 
$R$ remains silent if it cannot decode the received message correctly. 
$D$ and $E$ combine the two copies of the same signal received after 
two time slots to enhance their individual performance. There might be many possible diversity techniques that  
$D$ and $E$ can follow; we mainly focus on MRC and SC for this study. 
As MRC is the best diversity technique, implementation at $E$ can give the worst case secrecy analysis; 
on the other hand, if implemented at $D$, it can provide the best case secrecy given that the 
diversity techniques remain the same for $D$ and $E$ subsequently.

The channels are modeled as independent non-identical flat Rayleigh fading. The received SNR, 
$\gamma_{xy}$, of any arbitrary $x$-$y$ link from node $x$ to node $y$ can be 
expressed as
\begin{align}
\label{eq_1} 
\gamma_{xy} = \frac{P_{x}|h_{xy}|^2 }{N_{0_{y}}},
\end{align}
where $x$ and $y$ are from $\{S, R, D, E\}$ for any possible combination of 
$x$-$y$, $P_{x}$ is the power transmitted from node $x$, and $N_{0_{y}}$ is 
the noise variance of the additive white Gaussian noise (AWGN) at node $y$. 
As $|h_{xy}|$ is assumed to be following a Rayleigh distribution with average power 
unity, i.e., $\mathbb{E}[|h_{xy}|^2]=1$, $\gamma_{xy}$ is exponentially 
distributed with mean $1/\lambda_{xy}=P_{x}/N_{0_{y}}$. The CDF of 
$\gamma_{xy}$ can be written as 
\begin{align}
\label{eq_2} 
F_{\gamma_{xy}}(z)=1-\exp(-\lambda_{xy}z), ~~ z\geq 0. 
\end{align}
For notational simplicity, we further assume that the parameters of the 
$S$-$E$ and $R$-$E$ links are $\lambda_{xy}=\alpha_{se}$ and 
$\lambda_{xy}=\alpha_{re}$, respectively. The parameters of the other links, i.e., $S$-$D$, 
$S$-$R$, and $R$-$D$, are assumed to be $\lambda_{xy}=\beta_{sd}$, $\lambda_{xy}=\beta_{sr}$, 
and $\lambda_{xy}=\beta_{rd}$, respectively. 

The achievable secrecy rate is then given by \cite{wyner_wiretap, hellman_gaussian_wiretap}, 
\begin{align}
\label{eq_3}
C_{S}\triangleq{\frac{1}{2}\lb[\log_2\lb(\frac{1+\gamma_{M}}{1+\gamma_{E}}\rb)\rb]}^+,
\end{align} 
where $\gamma_{M}$ and $\gamma_{E}$ are the SNRs at $D$ and $E$, respectively. 
The term $1/2$ reflects the fact that two time slots are necessary for information transfer. 
The SOP is defined as the probability that the instantaneous 
secrecy capacity is less than a target secrecy rate, $R_s > 0$, as
\begin{align} 
\label{eq_4}
P_o\lb(R_s\rb)&=\mathbb{P}\left[C_{S} < R_s  \right] 
= \mathbb{P}\lb[\gamma_{M} < \rho\lb(1+\gamma_{E}\rb)-1 \rb]\nn\\&=\mathbb{E}_{\gamma_E}\lb[F_{\gamma_M}\lb(\rho\lb(1+\gamma_E\rb)-1\rb)\rb]
\end{align} 
where $\rho = 2^{2R_s}$.

\section{SOP of Various Combinations of Diversity Schemes}
\label{sec_sop}
This section evaluates the SOP under different combinations of diversity combining schemes 
considered at $D$ and $E$, when both $S$-$D$ and $S$-$E$ direct links exist. 
When $\gamma_{sr}>\gamma_{th}$, $R$ can correctly decode the source message; hence, 
both $R$-$E$ and $R$-$D$ links exist. Otherwise, these two links do not exist. 
If $D$ and $E$ perform MRC and SC, respectively, we denote the scheme 
as MRC-SC scheme. Similarly, we use MRC-MRC, SC-SC and SC-MRC schemes. In the following section, 
SOP is evaluated for two scenarios: when CSI is available at $S$ and $R$ transmitters 
and when it is not. Henceforth, we refer the transmitters of $S$ and $R$ simply as transmitters.
\subsection{CSI Unavailable at the Transmitters}
\label{subsec_unknown_csi_outage}
When knowledge of CSI is unavailable at the transmitters, they cannot adapt their rate 
according to the CSI. In this scenario, the SOPs are obtained for various combining schemes in the 
following subsections.  
\subsubsection{MRC-SC Scheme}
\label{sec_out_prob_mrc_sc}
% Let the instantaneous MRC, SC output SNRs at $D$ and $E$ be $\gamma_{M}$ and 
% $\gamma_{E}$ respectively. 
When $\gamma_{sr}>\gamma_{th}$, the output SNRs at $D$, $\gamma_{M}$, and $E$, $\gamma_{E}$, 
after MRC and SC diversity schemes, respectively, 
are \cite{Brennan_diversity, book_proakis}. 
% $\gamma_{M}$ and 
% $\gamma_{E}$ respectively $\gamma_{M}$ and $\gamma_{E}$ 
% are given as follows 
\begin{align}
\label{eq_9}
\gamma_{M}=\gamma_{sd}+\gamma_{rd},~~~
\gamma_{E}=\max\lb(\gamma_{se}, \gamma_{re}\rb). 
\end{align}
On the other hand, when $\gamma_{sr}<\gamma_{th}$, 
\begin{align}
\label{eq_10}
\gamma_{M}=\gamma_{sd},~~~
\gamma_{E}=\gamma_{se}.
\end{align}
% where and  are the output SNR at the $D$ and $E$ respectively.
The SOP of the system can be evaluated by finding the conditional SOP when $R$ correctly 
decodes the message and when it does not. From the theory of total probability, SOP 
can be expressed as
\begin{align}
\label{eq_12}
&P_o(R_s)=\mathbb{P}\lb[C_s<R_s|\gamma_{sr}>\gamma_{th}\rb]\mathbb{P}\lb[\gamma_{sr}>\gamma_{th}\rb] \nn\\
&+\mathbb{P}\lb[C_s<R_s|\gamma_{sr}<\gamma_{th}\rb]\mathbb{P}\lb[\gamma_{sr}<\gamma_{th}\rb] \nn \\
% \label{eq_12}
&=\int_o^\infty F_{\gamma_M}\lb(\rho\lb(1+x\rb)-1\rb|\gamma_{sr}>\gamma_{th})
f_{\gamma_E}(x|\gamma_{sr}>\gamma_{th})dx \nn\\&\times \lb[1-F_{\gamma_{sr}}\lb(\gamma_{th}\rb)\rb]  
\nn \\
% &+F_{\gamma_{sr}}\lb(\gamma_{th}\rb)\nn \\
&+\int_o^\infty F_{\gamma_M}\lb(\rho\lb(1+x\rb)-1\rb|\gamma_{sr}<\gamma_{th})
f_{\gamma_E}(x|\gamma_{sr}<\gamma_{th})dx \nn\\& \times F_{\gamma_{sr}}\lb(\gamma_{th}\rb). 
\end{align}
$F_{\gamma_{sr}}(\cdot)$ and $F_{\gamma_M}\lb(\cdot|\gamma_{sr} < \gamma_{th}\rb)$ 
can be obtained from (\ref{eq_2}). 
$F_{\gamma_M}\lb(\cdot|\gamma_{sr} \geq \gamma_{th}\rb)$, the CDF of 
the summation of two independent exponential distributions, can be obtained from \cite{sum_expo_mohamed_akkouchi}.
The PDF of the maximum of two arbitrary independent
exponentially distributed random variables with different parameters, $f_{\gamma_E}(\cdot|\gamma_{sr}<\gamma_{th})$, 
can also be easily obtained. 
% ,  and $f_{\gamma_E}(\cdot)$ are 
% evaluated , (\ref{eq_6}) and (\ref{eq_8}), respectively. 
% To obtain $F_{\gamma_M}\lb(\cdot|\gamma_{sr} \geq \gamma_{th}\rb)$, 
% $\lambda_1=\beta_{sd}$ and $\lambda_2=\beta_{rd}$ are substituted in 
% (\ref{eq_6}) and for $F_{\gamma_M}\lb(\cdot|\gamma_{sr} < \gamma_{th}\rb)$, 
% $\lambda_{xy}=\beta_{sd}$ is substituted in (\ref{eq_2}). 
The final expression 
is shown in (\ref{eq_19}), which is given in Table \ref{table_SOP_outg}.

\subsubsection{MRC-MRC Scheme}
\label{sec_out_prob_mrc_mrc}
Here we evaluate SOP when both $D$ and $E$ perform MRC. In the MRC-MRC scheme, 
the effective SNR at $D$ and $E$ is the sum of the link SNRs at those nodes. 
Under the condition $\gamma_{sr} \geq \gamma_{th}$, the received SNRs at the 
output of the MRC combiners at $D$ and $E$, respectively, are 
\begin{align}
\label{eq_16}
\gamma_{M}=\gamma_{sd}+\gamma_{rd}, ~~
\gamma_{E}=\gamma_{se}+\gamma_{re}.
\end{align} 
When $\gamma_{sr}<\gamma_{th}$, $\gamma_M$ and $\gamma_E$ are given in (\ref{eq_10}).
The SOP can be evaluated using (\ref{eq_12}), where, $F_{\gamma_M}(\cdot)$ and $f_{\gamma_E}(\cdot)$ 
are obtained from \cite{sum_expo_mohamed_akkouchi}. Finally, SOP is expressed as in 
(\ref{eq_20}) of Table \ref{table_SOP_outg}. 

\subsubsection{SC-SC Scheme}
\label{sec_out_prob_sc_sc}
Here we evaluate the SOP of the system when $D$ and $E$ both perform SC on the 
links received by them. Thus, when $\gamma_{sr} \geq \gamma_{th}$, the received SNRs 
at the output of the SC combiner at $D$ and $E$, respectively, are \cite{book_proakis}
\begin{align}
\label{eq_17}
\gamma_{M}=\max\lb(\gamma_{sd}, \gamma_{rd}\rb),~~ %\nn \\
\gamma_{E}=\max\lb(\gamma_{se}, \gamma_{re}\rb).
\end{align} 
When $\gamma_{sr}<\gamma_{th}$, $\gamma_M$ and $\gamma_E$ follow (\ref{eq_10}).
The SOP of the system is evaluated using (\ref{eq_12}),
% , where $F_{\gamma_{sr}}(\cdot)$, 
% $F_{\gamma_M}(\cdot)$, and $f_{\gamma_E}(\cdot)$ are obtained from (\ref{eq_2}), 
% (\ref{eq_7}), and (\ref{eq_8}), respectively. 
and is given in (\ref{eq_21}) of Table \ref{table_SOP_outg}.

\subsubsection{SC-MRC Scheme}
\label{sec_out_prob_sc_mrc}
We find the SOP of the SC-MRC combining scheme similarly to the previous sections. 
When $\gamma_{sr}\geq\gamma_{th}$, the received SNRs at the output of the SC and MRC 
combiner at $D$ and $E$, respectively, are \cite{book_proakis}
\begin{align}
\label{eq_18}
\gamma_{M}=\max\lb(\gamma_{sd}, \gamma_{rd}\rb),~~%\nn \\
\gamma_{E}=\gamma_{se}+\gamma_{re}.
\end{align} 
When $\gamma_{sr}<\gamma_{th}$, $\gamma_M$ and $\gamma_E$ can be obtained as in (\ref{eq_10}).
The SOP of the system can be evaluated from (\ref{eq_12}).
% , where $F_{\gamma_{sr}}(\cdot)$, 
% $F_{\gamma_M}(\cdot)$, and $f_{\gamma_E}(\cdot)$ are obtained from (\ref{eq_2}), (\ref{eq_7}), and 
% (\ref{eq_5}), respectively.
The final expression of SOP for SC-MRC scheme is given in (\ref{eq_22}) 
of Table \ref{table_SOP_outg}.

\subsection{CSI Available at the Transmitters}
\label{known_csi_outage}
This section evaluates SOP when complete CSI knowledge is available at the transmitters. 
As a result, $S$ can adapt its transmission rate to achieve positive secrecy. 
From the theorem of total probability, we can find the secrecy outage probability 
by calculating the  conditional SOP when $\gamma_{sr}\geq \gamma_{th}$ 
and $\gamma_{sr} <\gamma_{th}$. The conditional SOP must be obtained when 
$\gamma_M>\gamma_E$ for positive secrecy as
\begin{align}
\label{eq_23}
P_o(R_s)&=\mathbb{P}\lb[C_s<R_s\cap\gamma_M>\gamma_E|\gamma_{sr}>\gamma_{th}\rb]
\mathbb{P}\lb[\gamma_{sr}>\gamma_{th}\rb] \nn\\
&+\mathbb{P}\lb[C_s<R_s\cap\gamma_M>\gamma_E|\gamma_{sr}<\gamma_{th}\rb]
\mathbb{P}\lb[\gamma_{sr}<\gamma_{th}\rb].
\end{align}
% The quantity $\mathbb{P}\lb[C_s<R_s\cap\gamma_M>\gamma_E|\gamma_{sr}<\gamma_{th}\rb]$ 
$\mathbb{P}\lb[C_s<R_s\cap\gamma_M>\gamma_E|\gamma_{sr}>\gamma_{th}\rb]$ 
is evaluated as
\begin{align}
\label{eq_24}
&\mathbb{P}\lb[C_s<R_s\cap\gamma_M>\gamma_E|\gamma_{sr}>\gamma_{th}\rb]\nn\\
&=\mathbb{P}\lb[\gamma_E<\gamma_M<\rho(1+\gamma_E)-1|\gamma_{sr}>\gamma_{th}\rb] \nn \\
&=\int_0^\infty\int_y^{\rho(1+y)-1} f_{\gamma_M}(x) f_{\gamma_E}(y)dx dy \nn\\
&=\int_0^\infty[F_{\gamma_M}\lb(\rho\lb(1+y\rb)-1\rb)-F_{\gamma_M}(y)]f_{\gamma_E}(y) dy .
\end{align}

When knowledge of CSI is available at the transmitters, the SOP
of MRC-SC, MRC-MRC, SC-SC, and SC-MRC schemes are directly obtained, as given 
in Table \ref{table_SOP_csi}.

\section{ESR of Various Combinations of Diversity Schemes}
\label{sec_erg_sec_rate}
We find the ESR, $\bar{C}_{S}$, under two scenarios, i.e.,  
when complete CSI knowledge is available at the transmitters 
and when the CSI knowledge is unavailable.
$\bar{C}_{S}$ can be expressed as \cite{Olabiyi_Sec}
\begin{align}
\label{eq_29}
\bar{C}_{S}&=\bar{C}_{S}({\gamma_{sr}\geq\gamma_{th}})\mathbb{P}\lb[\gamma_{sr}\geq\gamma_{th}\rb] \nn\\
&+\bar{C}_{S}({\gamma_{sr}<\gamma_{th}})\mathbb{P}\lb[\gamma_{sr}<\gamma_{th}\rb] \nn \\
&=\bar{C}_{S}({\gamma_{sr}\geq\gamma_{th}})\lb(1-\mathbb{P}\lb[\gamma_{sr}<\gamma_{th}\rb]\rb)\nn\\
&+\bar{C}_{S}({\gamma_{sr}<\gamma_{th}})\mathbb{P}\lb[\gamma_{sr}<\gamma_{th}\rb],
\end{align}
where $\bar{C}_{S}({\gamma_{sr}\geq\gamma_{th}})$ is the conditional ESR when 
$\gamma_{sr}\geq\gamma_{th}$ and, similarly, $\bar{C}_{S}({\gamma_{sr}<\gamma_{th}})$ is the 
conditional ESR when $\gamma_{sr}<\gamma_{th}$.  
Further, $\mathbb{P}\lb[\gamma_{sr}<\gamma_{th}\rb]$ can be found from (\ref{eq_2}).

\subsection{CSI Unavailable at the Transmitters}
\label{unknown_csi}
In (\ref{eq_29}),
$\bar{C}_{S}({\gamma_{sr}\geq\gamma_{th}})$ can be evaluated from (\ref{eq_3}) as
\begin{align}
\label{eq_30}
&\bar{C}_{S}({\gamma_{sr}\geq\gamma_{th}})\nn\\
&=\frac{1}{2\ln2}\int_0^\infty \int_0^\infty{\ln\lb[\frac{1+x}{1+y}\rb]}
f_{\gamma_{M}}(x|{\gamma_{sr}\geq\gamma_{th}}) \nn\\
&\times f_{\gamma_{E}}(y|{\gamma_{sr}\geq\gamma_{th}}) dxdy\nn \\
&=\frac{1}{2\ln2} \lb[\bar{I}_M({\gamma_{sr}\geq\gamma_{th}})-\bar{I}_E({\gamma_{sr}\geq\gamma_{th}})\rb],
\end{align}
where $\bar{I}_M({\gamma_{sr}\geq\gamma_{th}})$ and $\bar{I}_E({\gamma_{sr}\geq\gamma_{th}})$ are 
expressed respectively as
\begin{align}
\label{eq_31}
\bar{I}_M(\gamma_{sr}\geq\gamma_{th}) 
&=\int_0^\infty \ln\lb(1+x\rb)f_{\gamma_M}(x|{\gamma_{sr}\geq\gamma_{th}})dx, \\
% \end{align}
% and
% \begin{align}
\label{eq_32}
\bar{I}_E(\gamma_{sr}\geq\gamma_{th})
&=\int_0^\infty \ln\lb(1+y\rb)f_{\gamma_E}(y|{\gamma_{sr}\geq\gamma_{th}})dy.
\end{align}  
The integrals in (\ref{eq_31}) and (\ref{eq_32}) can be evaluated over $x$ and $y$ 
separately over entire range, as the knowledge of CSI is not available to 
impose any limit other than zero to infinity on the respective integration. 

Further, $\bar{C}_{S}\lb({\gamma_{sr}<\gamma_{th}}\rb)$ can be evaluated following 
a similar way from (\ref{eq_30}) to (\ref{eq_32}). Substituting 
$F_{\gamma_M}(\cdot)$ and $f_{\gamma_E}(\cdot)$ for the various diversity combining techniques 
in (\ref{eq_30}), ESR can be derived and results are listed in Table \ref{table_sec_rate} from 
(\ref{eq_36}) to (\ref{eq_42}). 

For the final derivation of (\ref{eq_29}), we have used the integral solution of the form 
\cite[(2.6.23.5)]{book_Prudnikov_v1}
\begin{align}
\label{eq_A36}
\int_0^\infty e^{-px} \ln \lb(a+bx\rb)  dx = \frac{1}{p} \lb[ \ln a - e^{\frac{ap}{b}} 
\operatorname{Ei}{ \lb(-\frac{ap}{b}\rb)}\rb].
\end{align}
In (\ref{eq_A36}), $\operatorname{Re} (p)  > 0$, $|\operatorname{arg} (\frac{a}{b})|<\pi$,  $\operatorname{Re}(\cdot)$ 
is the real part of its argument, $\operatorname{arg}(\cdot)$ represents the argument of a complex 
quantity, and the exponential integral function $\operatorname{Ei}(\cdot)$ is given by
\begin{align}
\label{eq_A37}
\operatorname{Ei}(x)=\int_{-\infty}^{x}\frac{e^t}{t} dt. 
\end{align}

\subsection{CSI Available at the Transmitters}
\label{known_csi}
If the CSI information is available while evaluating (\ref{eq_29}), we can adapt the transmission only when $\gamma_M>\gamma_E$. 
As in (\ref{eq_30}), $\bar{C}_{S}({\gamma_{sr}\geq\gamma_{th}})$ can be evaluated as
\begin{align}
\label{eq_33}
\bar{C}_{S}({\gamma_{sr}\geq\gamma_{th}})
% \nn\\
% &=\frac{1}{2\ln2}\int_0^\infty \int_0^x {\ln\lb[\frac{1+x}{1+y}\rb]} f_{\gamma_{M}}(x|{\gamma_{sr}\geq\gamma_{th}})\nn\\
% &\times f_{\gamma_{E}}(y|{\gamma_{sr}\geq\gamma_{th}}) dxdy\nn \\
&=\frac{1}{2\ln2}\lb[\bar{I}_M({\gamma_{sr}\geq\gamma_{th}})-\bar{I}_E({\gamma_{sr}\geq\gamma_{th}})\rb],
\end{align}
% where $\bar{I}_M({\gamma_{sr}\geq\gamma_{th}})$ is $\bar{I}_M$ conditioned on $\gamma_{sr}\geq\gamma_{th}$
% and $\bar{I}_E({\gamma_{sr}\geq\gamma_{th}})$  is $\bar{I}_E$ conditioned on $\gamma_{sr}\geq\gamma_{th}$. 
where $\bar{I}_M({\gamma_{sr}\geq\gamma_{th}})$ and $\bar{I}_E({\gamma_{sr}\geq\gamma_{th}})$ can be 
expressed respectively  as
\begin{align}
\label{eq_34}
&\bar{I}_M(\gamma_{sr}\geq\gamma_{th})\nn\\&=\int_0^\infty \int_0^x \ln\lb(1+x\rb) 
f_{\gamma_E}(y|\gamma_{sr}\geq\gamma_{th})f_{\gamma_M}(x|{\gamma_{sr}\geq\gamma_{th}}) dydx \\
\label{eq_35}
&\bar{I}_E(\gamma_{sr}\geq\gamma_{th})\nn\\
&=\int_0^\infty \int_0^x \ln\lb(1+y\rb)f_{\gamma_E}(y|\gamma_{sr}\geq\gamma_{th}) 
f_{\gamma_M}(x|\gamma_{sr}\geq\gamma_{th}) dydx \nn \\
% &=\int_0^\infty \int_y^\infty \ln(1+y)f_{\gamma_E}(y|\gamma_{sr}\geq\gamma_{th}) 
% f_{\gamma_M}(x|\gamma_{sr}\geq\gamma_{th})dxdy \nn \\
&=\int_0^\infty \ln(1+y)F_{\gamma_{M}}^c(y|\gamma_{sr}\geq\gamma_{th}) 
f_{\gamma_{E}}(y|\gamma_{sr}\geq\gamma_{th}) dy, 
\end{align}  
where $F_{\gamma_{M}}^c(\cdot)= 1-F_{\gamma_{M}}(\cdot)$. Unlike (\ref{eq_31}) and (\ref{eq_32}), it can be noticed that (\ref{eq_34}) and (\ref{eq_35}) have 
an upper limit on $\gamma_E$ to make sure that  $\gamma_E<\gamma_M$, following the knowledge of CSI. 
Further, $\bar{C}_{S}\lb({\gamma_{sr}<\gamma_{th}}\rb)$ can be evaluated following a similar way as in 
(\ref{eq_33})-(\ref{eq_35}).
By replacing $F_{\gamma_M}(\cdot)$ and $f_{\gamma_E}(\cdot)$ for different combining schemes in (\ref{eq_34})-(\ref{eq_35}), 
we can find the ESR of the corresponding systems. 
% When direct link $S$-$D$ and $S$-$E$ are 
% absent, (\ref{eq_23}) is evaluated to $\bar{C}_s^{NOD}$, when MRC and SC is performed at $E$ (\ref{eq_23}) is evaluated 
% to $\bar{C}_s^{MRC}$ and $\bar{C}_s^{SC}$  respectively. 
Finally, the results are provided in Table \ref{table_sec_rate} from 
(\ref{eq_37}) to (\ref{eq_43}).

% A factor of $1/2\ln2$ is multiplied 
% with each equation in Table \ref{table_sec_rate} but is not shown in table for brevity. 

\section{Asymptotic Analysis}
\label{sec_asymp_ana}
We are interested in finding the asymptotic expressions of $P_o(R_s)$ for the following two cases: 
\textit{A}) when $S$-$R$ and $R$-$D$ links average SNRs tends to infinity. 
This is the balanced case; and \textit{B}) 
when the average SNR of either $S$-$R$ or $R$-$D$ link tends to infinity 
while that of the other link is fixed. This is the unbalanced case. 
The scenario of unbalanced links might arise due to unequal transmit power at $S$ or $R$. It can also arise 
if $R$ is not placed at an identical distance from $S$ and $D$, with identical $S$-$R$ and $R$-$D$ links.

\subsection{Balanced Case}
\label{asymp_balance}
The asymptotic behaviour in the balanced case can provide a limiting behavior of $P_o(R_s)$ when both dual hop links are 
quite strong when compared to the direct links to the $D$ and $E$. We obtain the asymptotic expression by 
setting $1/\beta_{sr}=1/\beta_{rd}=1/\beta\rightarrow  \infty$.
Under such condition, from (\ref{eq_19}) and after some manipulations, 
the  asymptotic SOP of the MRC-SC scheme when CSI is unavailable can be expressed as in (\ref{eq_44}).
As MRC is a superior diversity technique than SC, MRC applied at $D$ and SC applied at $E$ will 
provide best secrecy performance. Likewise, SC-MRC will provide the worst secrecy. As a result, the performances of 
all combinations of these two diversity schemes will lie in between the MRC-SC and SC-MRC schemes. On the other hand, 
in the MRC-MRC scheme, $D$ and $E$ both utilize best possible diversity scheme. Hence, 
we have derived the asymptotes of these three schemes for the cases when CSI is available or not 
and the results are given in Table \ref{talbe_SOP_bal}. Asymptotic SOP is inversely proportional to the SNR of the 
balanced links, hence, it can be understood that the secrecy of the system can be 
improved by improving the balanced dual-hop link.  

Secrecy of two particular cases: i) when $R$ can always decode the message properly, as it is generally assumed in the 
literature; and ii) the traditional wiretap channel, which can be obtained 
by simply choosing $\gamma_{th}$ properly in our proposed threshold-selection 
relaying technique. 
% The assumption that $R$ can always decode the source message correctly can be realized by simply 
% setting $\gamma_{th} \rightarrow 0$. On the other hand, if $\gamma_{th}$ is infinitely high, $R$ will not be able to transmit 
% the message and the system reduces to a wiretap channel. 
Wiretap channel SOP 
is the same irrespective of combining schemes; on the contrary, when $\gamma_{th} \rightarrow 0$, 
different combining schemes provide different SOPs. We obtained the asymptotic SOPs under these two limiting 
cases for the {MRC-MRC}\footnote{Only MRC-MRC is shown for the illustration purpose. 
The asymptotic SOPs for the other combining schemes can be derived similarly having same analytical behaviour.} 
scheme when CSI is unavailable.
% and plotted in Fig. \ref{FIG_4}. 
The asymptotic expression of SOP for $\gamma_{th} \rightarrow 0$ 
can be evaluated from (\ref{eq_20}) as 
\begin{align}
\label{eq_45}
P_o^{AS}(R_s)&=1 
-\frac{\alpha_{se}\alpha_{re}}{\lb(\beta_{sd}-\beta_{rd}\rb)}
\lb[\frac{\beta_{sd}e^{-\beta_{rd}(\rho-1)}}
{\lb(\alpha_{se}+\rho\beta_{rd}\rb)\lb(\alpha_{re}+\rho\beta_{rd}\rb)} \rb.\nn\\
&\lb.-\frac{\beta_{rd}e^{-\beta_{sd}(\rho-1)}}
{\lb(\alpha_{se}+\rho\beta_{sd}\rb)\lb(\alpha_{re}+\rho\beta_{sd}\rb)}\rb].
\end{align}
The corresponding expression for ESR is evaluated from (\ref{eq_36_1}) as
\begin{align}
\label{eq_59}
\bar{C}_s^{AS}&=\frac{1}{\beta_{sd}-\beta_{rd}}
\lb(\beta_{rd}e^{\beta_{sd}}\operatorname{Ei}\lb(-\beta_{sd}\rb) 
-\beta_{sd}e^{\beta_{rd}}\operatorname{Ei}\lb(-\beta_{rd}\rb) \rb) \nn \\
&-\frac{1}{\alpha_{se}-\alpha_{re}}\lb(\alpha_{re}e^{\alpha_{se}}\operatorname{Ei}\lb(-\alpha_{se}\rb)
-\alpha_{se}e^{\alpha_{re}}\operatorname{Ei}\lb(-\alpha_{re}\rb)\rb).
\end{align}
It can be noticed that both (\ref{eq_45}) and (\ref{eq_59}) are independent of $\beta_{sr}$. This is intuitive as
$R$ is going to decode correctly irrespective of the $S$-$R$ link quality.

When $\gamma_{th} \rightarrow \infty$, the asymptotic SOP can be expressed by
\begin{align}
\label{eq_46}
 P_o^{AS}(R_s)= 1-\frac{\alpha_{se}e^{-\beta_{sd}(\rho-1)}}{\alpha_{se}+\rho\beta_{sd}}.
\end{align}
It can be readily observed that (\ref{eq_46}) is the SOP of the wiretap channel \cite{McLaughlin_wireless_info_theo_sec}.
The threshold SNR at the relay as the condition for the correct detection actually generalizes the performance with 
perfect decoding and wiretap channel. The corresponding ESR can be found in \cite{Olabiyi_Sec}, hence, not evaluated.

\subsection{Unbalanced Case}
\label{asymp_unbalance_rd}
Unbalance in the system means that the $S$-$R$ or $R$-$D$ links have different average SNRs. This can 
arise if either one of the $S$-$R$ or $R$-$D$ links is closely spaced when compared to the other links. 
Unbalance is studied in the following two cases. 
In Case I, we study the behavior of the SOP keeping the average SNR of the $S$-$R$ link fixed 
and asymptotically increasing the average SNR of the $R$-$D$ link. 
In Case II, we study the behavior of SOP keeping the average SNR of the $R$-$D$ link fixed 
and asymptotically increasing the average SNR of the $S$-$R$ link.

Asymptotic SOPs are evaluated similarly to the balanced case for the MRC-SC and SC-MRC schemes 
along with the MRC-MRC scheme. 
These are evaluated from (\ref{eq_19}), (\ref{eq_20}) and (\ref{eq_22}) when CSI is unavailable, and 
from (\ref{eq_25}), (\ref{eq_26}) and (\ref{eq_28}) when CSI is available, respectively. 
In Case I, when $1/\beta_{sr}$ is fixed and $1/\beta_{rd}=1/\beta\rightarrow  \infty$,  
or in Case II, when $1/\beta_{rd}$ is fixed and $1/\beta_{sr}=1/\beta\rightarrow  \infty$,  
the asymptotic SOPs can be expressed as a summation of a constant quantity independent of SNR ($1/\beta$)
and an asymptotically varying term, which depends inversely on $1/\beta$. This can be seen from 
Tables \ref{table_SOP_unbal_I} and \ref{table_SOP_unbal_II} for Cases I and II, 
respectively. At low SNR, the varying term dominates; however, it vanishes at high SNR. 
% We can see that constant terms are the same for Case I in Table \ref{table_SOP_unbal_I}, 
% on the contrary, they are different for Case II in Table \ref{table_SOP_unbal_II}.
It is understood from the asymptotic analysis in the unbalanced cases that 
SOP saturates to a certain value gradually with the increase in the SNR of the unbalanced link. 
The weak link can be a bottleneck to improve overall security in this kind of dual hop 
cooperative systems with threshold-selection DF relay.

The asymptotic ESR in Case II is derived for the 
MRC-SC\footnote{Only MRC-SC is shown for the illustration  purpose when CSI is available. 
Derivation for the other combining schemes when CSI is unavailable
is identical having similar analytical behaviour.} scheme when CSI is available 
from (\ref{eq_36}), which is 
displayed in (\ref{eq_39}).
% , and plotted in Fig. \ref{FIG_7}.
% \begin{figure*}
% [!t]
\begin{align}
\label{eq_39}
\bar{C}_s&=\frac{1}{2\ln2\lb(\beta_{sd}-\beta_{rd}\rb)}
\lb[\beta_{rd}\lb(e^{\beta_{sd}}\operatorname{Ei}\lb(-\beta_{sd}\rb)\rb.\rb.\nn\\
&\lb.\lb.
-e^{\beta_{sd}+\alpha_{se}}\operatorname{Ei}\lb(-\beta_{sd}-\alpha_{se}\rb)
-e^{\beta_{sd}+\alpha_{re}}\operatorname{Ei}\lb(-\beta_{sd}-\alpha_{re}\rb) 
\rb.\rb.\nn\\
&\lb.\lb.+e^{\beta_{sd}+\alpha_{se}+\alpha_{re}}\operatorname{Ei}\lb(-\beta_{sd}-\alpha_{se}-\alpha_{re}\rb)\rb) \rb.\nn\\
&\lb.-\beta_{sd}\lb(e^{\beta_{rd}}\operatorname{Ei}\lb(-\beta_{rd}\rb)
-e^{\beta_{rd}+\alpha_{se}}\operatorname{Ei}\lb(-\beta_{rd}-\alpha_{se}\rb)\rb.\rb.\nn\\
&\lb.\lb.-e^{\beta_{rd}+\alpha_{re}}\operatorname{Ei}\lb(-\beta_{rd}-\alpha_{re}\rb)
\rb.\rb.\nn\\
&\lb.\lb.+e^{\beta_{rd}+\alpha_{se}+\alpha_{re}}\operatorname{Ei}\lb(-\beta_{rd}-\alpha_{se}-\alpha_{re}\rb)\rb) \rb].
\end{align}
% \hrule
% \end{figure*}
\section{Numerical and Simulation Results} 
\label{sec_results} 
This section describes the numerical results, validated by simulations. 
% All the link SNRs, threshold are in dB unit. 
Without loss of generality, results are obtained assuming that all nodes are affected by the same noise power, $N_{0}$. 
In the figures, CSI indicates that results are obtained when CSI is available and NOCSI indicates that results are obtained 
when CSI is unavailable.  The unit of $R_s$ is bits per channel use (bpcu). 
Unless otherwise specified, simulation parameters are:
$\gamma_{th} = 3$ dB, $1/\alpha_{se} = 0$ dB, $1/\alpha_{re} = 3$ dB, $1/\beta_{sd} = 3$ dB, 
$1/\beta_{rd} = 3$ dB, $R_s = 1 $ bpcu. The  blue colour shows results for
NOCSI, whereas, red or black represents results for CSI. 

\begin{figure} 
% [H]
\centering
% \vspace*{-3.5cm}
\includegraphics[width=0.5\textwidth]{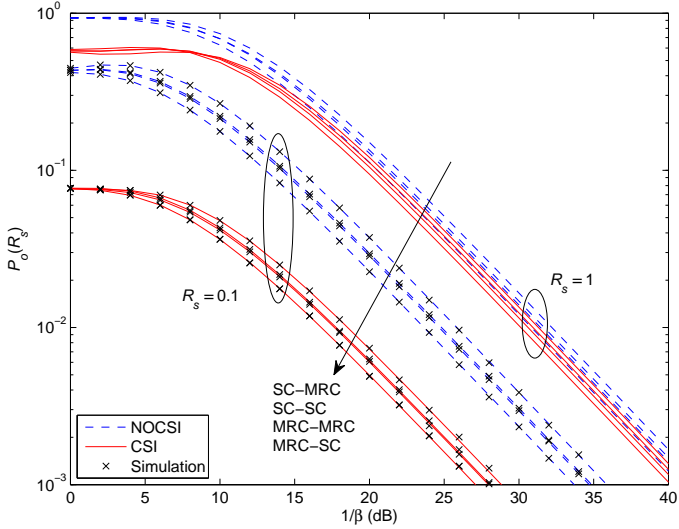}
% \vspace*{.3cm}
\caption{
SOP of diversity combining schemes for CSI 
and NOCSI in the balanced case with  
$\gamma_{th} = 3$ dB, $1/\alpha_{se} = 0$ dB, $1/\alpha_{re} = 3$ dB,  
$1/\beta_{sd} = 3$ dB, and $R_s = 0.1, 1$ bpcu.
% $1/\beta_{sr} = 1/\beta_{rd}$.
}
\label{FIG_2}
\vspace*{-0.5cm}
\end{figure}

\textbf{Effect of $R_s$ on SOP}: Fig. \ref{FIG_2} 
shows the SOP versus average SNR for the diversity combining 
schemes evaluated in Section \ref{sec_sop} for the balanced case.
% using (\ref{eq_19}), (\ref{eq_20}), (\ref{eq_21}) and (\ref{eq_22}). 
% when $1/\beta_{sd} = 3$ dB, $1/\alpha_{se} = 0$ dB, $1/\alpha_{re} = 3$ dB, and $\gamma_{th} = 3$. 
SOPs are compared for NOCSI ((\ref{eq_19})-(\ref{eq_22})) and CSI ((\ref{eq_25})-(\ref{eq_28})) for different $R_s = 0.1$ 
and $1$ bpcu.
It can be seen that the order of the performance of SOP from the best to the worst is: 
MRC-SC, MRC-MRC, SC-SC, and SC-MRC, respectively. MRC is the optimal combining technique whose performance is 
better than SC; hence, the combination of MRC at the $D$ and SC at the $E$ 
yields the best SOP performance. It is intuitive to observe that SOP is higher when $R_s$ is higher. 
As expected, the availability of CSI can provide a better performance when
compared to NOCSI, however, the degree of improvement is higher when $R_s$ is lower.
When $R_s$ is higher, SOP itself tends to get higher; hence, knowledge of CSI cannot significantly overcome the SOP 
induced by high $R_s$. Simulation results are shown only for low $R_s$ for better clarity; 
these match exactly with the analytical results. 

\begin{figure} 
% [H]
\centering
% \vspace*{-3.5cm}
\includegraphics[width=0.5\textwidth]{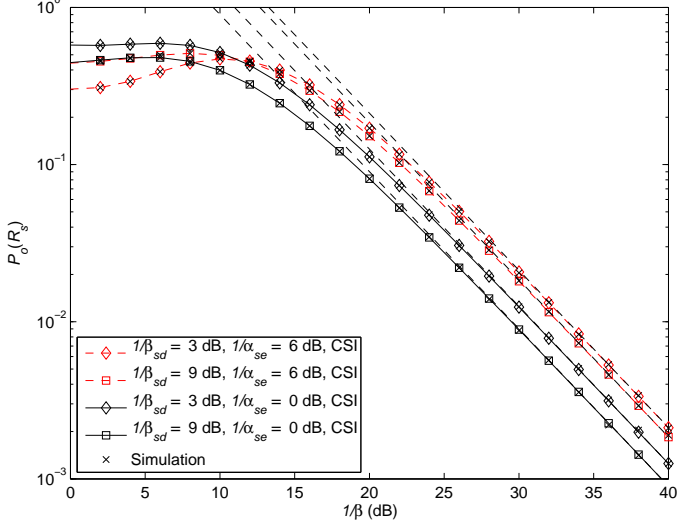}
% \vspace*{.3cm}
\caption{
SOP of the MRC-MRC scheme for CSI in the balanced case with 
$\gamma_{th} = 3$ dB, $1/\alpha_{se} = 0, 6$ dB, $1/\alpha_{re} = 3$ dB, 
$1/\beta_{sd} = 9, 3$ dB, and $R_s = 1 $ bpcu. Straight dashed lines represent asymptotes.
% $1/\beta_{sr} = 1/\beta_{rd}$.
} 
\label{FIG_3}
\vspace{-0.5cm}
\end{figure}

\textbf{SOP by improving main link quality for a given eavesdropper link quality:}
Fig. \ref{FIG_3} depicts the SOP versus average SNR for the 
MRC-MRC\footnote{Only the SOP of MRC-MRC scheme is shown when CSI is available to maintain better clarity; however, 
conclusions from observations are applicable in general irrespective of 
combining schemes and CSI availability.} scheme when CSI 
is available in the balanced case. The figure is obtained by improving $1/\beta_{sd}$ 
from $3$ dB to $9$ dB for a given $1/\alpha_{se}=0$ or $6$ dB with the help of  (\ref{eq_26}).
The asymptotes are plotted using dashed straight lines with the help of (\ref{eq_47_1}). 
It can be observed that an increase of 6 dB in the $S$-$D$ channel quality improves SOP for 
a given eavesdropper link quality. 
However, it is interesting to note that the amount of improvement is higher when the eavesdropper channel quality is 
lower, i.e., $1/\alpha_{se}=0$ dB. This can be easily understood by comparing the gap between two asymptotes (dashed lines) 
corresponding to $1/\alpha_{se}=0$ dB and $1/\alpha_{se}=6$ dB.
Intuitively, it turns out to be difficult to improve secrecy when secrecy itself is low 
due to good eavesdropper channel quality.   
% Some general observations 
% like improvement in the $S$-$D$  link quality improves the SOP or improvement in the $S$-$E$ link 
% quality worsen the SOP can also be observed.

\begin{figure}
% [t]
\centering
% \vspace*{-3.5cm}
\includegraphics[width=0.55\textwidth] {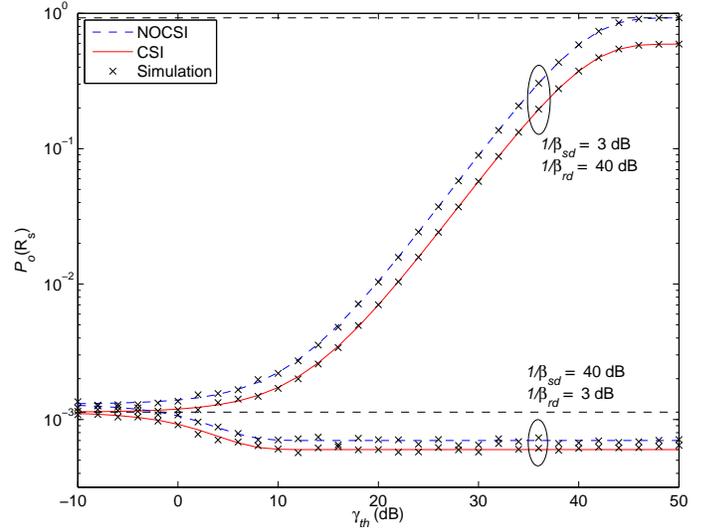}
% \vspace*{4.5cm}
\caption{SOP of the MRC-MRC scheme versus $\gamma_{th}$ for CSI and NOCSI in the balanced case with 
% $\gamma_{th} = -10:2:50 $dB, 
$1/\alpha_{se} = 0$ dB, $1/\alpha_{re} = 3$ dB, 
$1/\beta_{sd} = 40,~ 3$ dB, $1/\beta_{rd} = 3, ~40$ dB, and $R_s = 1 $ bpcu. Horizontal dashed 
lines represent saturation levels when $\gamma_{th}\rightarrow \infty$ and $\gamma_{th}\rightarrow 0$, 
respectively.}
\label{FIG_4}
\vspace{-0.5cm}
\end{figure}

\textbf{Effect of the $S$-$D$ and $R$-$D$ link qualities
on SOP}: Fig. \ref{FIG_4} depicts the SOP 
performance versus $\gamma_{th}$ of the MRC-MRC\footnotemark
\footnotetext{Only the SOP of MRC-MRC scheme is shown for better clarity as illustration purpose.} 
scheme in the balanced case for CSI and NOCSI. 
The SOP is obtained when the $S$-$D$ link quality is very high as compared to the $R$-$D$ link quality, i.e., 
$1/\beta_{sd}=40$ dB and $1/\beta_{rd}=3$ dB and vice versa.
It can be observed that as $\gamma_{th}$ increases, SOP increases 
when the $S$-$D$ link quality is very low when compared to the $R$-$D$ link quality; 
on the other hand, it decreases when the $S$-$D$ link quality is very high when compared to the $R$-$D$ link quality.
% depending on the $S$-$D$ and $S$-$E$ link quality.
% If $\gamma_{th}$ increases, the probability of correct decision at $R$ decreases which prevents 
% $R$ to transmit its decision. 
An increase in $\gamma_{th}$ decreases the probability of relaying,
% ; as a result, % relayed transmission to $D$ can be interrupted. 
and hence, if the $R$-$D$ link quality is far better than the $S$-$D$ link, SNR at $D$ can be decreased
significantly and SOP can be decreased. 
On the other hand, an increase in $\gamma_{th}$ can decrease the transmission towards $E$ as well, 
hence, when the $S$-$D$ link quality is far better than the $R$-$D$ link quality, SNR at 
$D$ remains nearly unchanged, however, SNR at $E$ decreases. As a result, SOP can be decreased. 

\textbf{Asymptotic behaviour of SOP with respect to $\gamma_{th}$}:
In Fig. \ref{FIG_4}, it can also be seen that 
SOP saturates to a certain value if $\gamma_{th}$ increased to infinity or decreased to zero.
If $\gamma_{th}\rightarrow  \infty$, there can be no transmission from $R$; 
hence, SOP saturates to a fixed value irrespective of the combining schemes. 
This is shown by the dashed line on the top of the figure which is the SOP of the 
wiretap channel evaluated in (\ref{eq_46}). When $\gamma_{th}\rightarrow 0$, i.e., $R$ 
always decodes the message correctly, the SOP also saturates to a fixed value shown by 
the dashed line at the bottom of the figure, evaluated from (\ref{eq_45}).

\begin{figure}
%  [t]
\centering
% \vspace*{-3.5cm}
\includegraphics[width=0.5\textwidth] {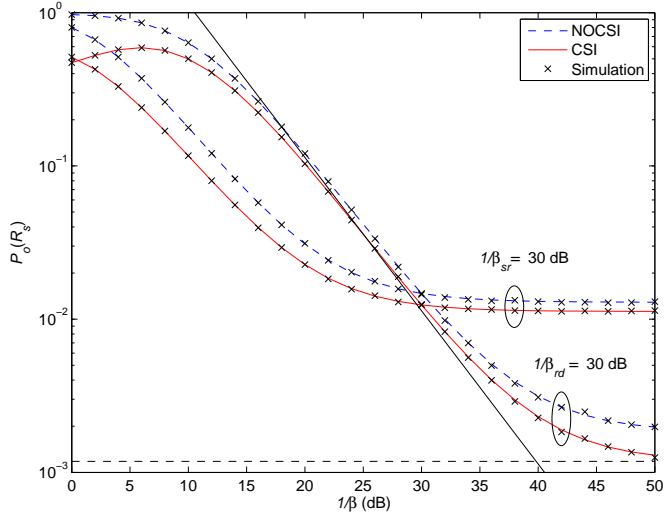}
% \vspace*{4.5cm}
\caption{SOP of the MRC-MRC scheme in the unbalanced case with 
$\gamma_{th} = 3$ dB, $1/\alpha_{se} = 0$ dB, $1/\alpha_{re} = 3$ dB, 
$1/\beta_{sd} = 3$ dB, $1/\beta_{sr} = 30$ dB (Case I) or $1/\beta_{rd} = 30$ dB (Case II), and 
$R_s = 1 $ bpcu. Solid straight line represents asymptote and the dashed straight 
line represents saturation level when $1/\beta_{sr}\rightarrow \infty$.}
\label{FIG_5}
\vspace{-0.5cm}
\end{figure}

\textbf{Effect of unbalance in the dual-hop links on SOP}: Fig. \ref{FIG_5} plots the SOP versus average SNR of the 
MRC-MRC\footnotemark[\value{footnote}] scheme when there is unbalance in the 
dual-hop links for CSI and NOCSI. 
The SOP is obtained by increasing $1/\beta_{rd}$ for a fixed  
$1/\beta_{sr} = 30$ dB in the unbalanced Case I and also increasing $1/\beta_{sr}$ for a given 
$1/\beta_{rd} = 30$ in the unbalanced Case II. The $x$-axis represents $1/\beta_{rd}$ in 
Case I and $1/\beta_{sr}$ in Case II.
It can be seen that for a given $1/\beta_{sr}$ or $1/\beta_{rd}$, SOP saturates to a particular value. 
The saturation value is basically the constant term shown in Tables \ref{table_SOP_unbal_I} 
and \ref{table_SOP_unbal_II}. The constant term of MRC-MRC scheme is shown with a horizontal dashed line 
and the corresponding asymptotically varying term by a solid straight 
line when CSI is available for Case II,  with the help of (\ref{eq_56_X}).

Careful observation into Table \ref{table_SOP_unbal_I} reveals that the constant terms for all 
diversity schemes are the same in Case I for CSI. 
This is also true for NOCSI, however, the constant terms for CSI and NOCSI are different.  
On the contrary, Case II has different constant terms for CSI and NOCSI
% whether CSI in available or unavailable 
in Table \ref{table_SOP_unbal_II}. 
% This can be observed by looking at the equations from (\ref{eq_54}) to (\ref{eq_57_1}).
In Case I, where $1/\beta_{sr}$ is a fixed value and $1/\beta_{rd}$ tends to infinity, 
the probability that the received SNR at $R$ exceeds $\gamma_{th}$ is 
the same irrespective of diversity schemes. Further, as the $R$-$D$ 
link SNR is very high, all diversity schemes tend to produce the same performance. 
Hence, the constant terms saturate to the same 
value depending on the $S$-$R$ link SNR.
On the other hand, in Case II where $1/\beta_{rd}$ is a fixed value and $1/\beta_{sr}$ 
tends to infinity, though SNR of the $S$-$R$ link always exceeds $\gamma_{th}$, due to fixed 
$R$-$D$ link SNR, performance saturates to different values for different diversity schemes. 
In conclusion, it is clear that unbalance in the dual hop link can serve as a bottleneck to the SOP 
performance, and these two unbalances are not identical. 
The performance cannot be improved even if the average SNR of the
$R$-$D$ link is increased to infinity keeping the $S$-$R$ link SNR fixed or vice versa.

\begin{figure}
% [H]
\centering
% \vspace*{-3.5cm}
\includegraphics[width=0.5\textwidth]{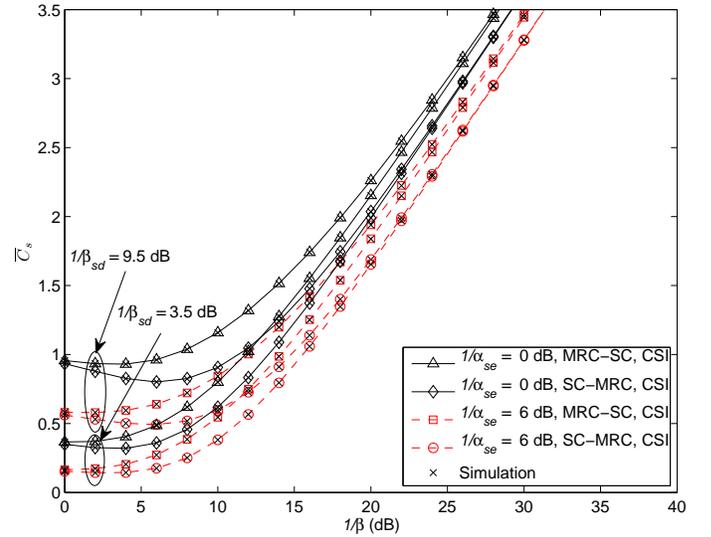}
% \vspace*{4.5cm}
\caption{
ESR of MRC-SC and SC-MRC schemes under CSI and NOCSI assumption in the balanced case with
$\gamma_{th} = 3$ dB, $1/\alpha_{se} = 3$ dB, $1/\alpha_{re} = 3.5$ dB, and 
$1/\beta_{sd} = 9.5, 3.5$ dB.  
% $1/\beta_{sr} = 1/\beta_{rd}$, SC-MRC, MRC,SC
} 
\label{FIG_6}
\vspace{0.5cm}
\end{figure}

\textbf{Effects of the $S$-$E$ and $S$-$D$ link qualities on ESR}:
Fig. \ref{FIG_6} shows the ESR versus average SNR for the MRC-SC and SC-MRC schemes when 
CSI is available in the balanced case. 
Results are obtained by increasing $1/\alpha_{se}$ from $0$ dB to $6$ dB when $1/\beta_{sd} = 9.5$ and $3.5$ dB 
at $1/\alpha_{re}=3.5$ dB. The results are evaluated using (\ref{eq_37}) and  (\ref{eq_43}) 
for MRC-SC and SC-MRC schemes, respectively.
It can be observed that for a given diversity scheme,
% for example, let us say MRC-SC scheme, 
% as the average SNR improves, 
ESR for different $S$-$D$ link quality
% value, i.e., 
% $1/\beta_{sd}=9.5$ dB and $1/\beta_{sd}=3.5$ dB, 
gradually merges with each other if the $S$-$E$ link quality is the same. 
On the contrary, ESR is different for different 
$S$-$E$ link qualities,  even if the $S$-$D$ link quality is the same.
% This can be easily followed by observing that black and red curves 
% from the same group of $1/\beta_{sd}$ do not match for a given diversity scheme. 
This suggests that the secrecy is more sensitive to changes in $S$-$E$ link quality 
than changes in the $S$-$D$ link quality. Further, some general observations can be made as 
ESR improves with an increase in the $S$-$D$ link quality and decreases with the 
increase in the $S$-$E$ link quality.

\begin{figure} 
% [H]
\centering
% \vspace*{-3.5cm}
\includegraphics[width=0.5\textwidth]{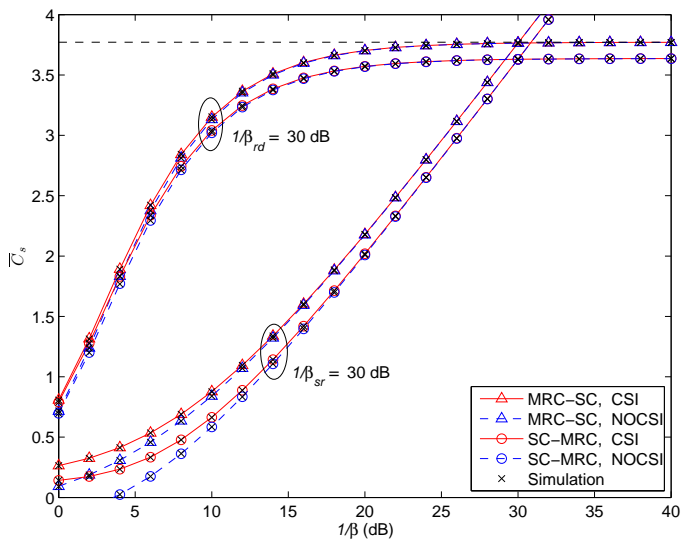}
% \vspace*{.3cm}
\caption{
ESR of MRC-SC and SC-MRC schemes under CSI and NOCSI assumption in the unbalanced case with 
$\gamma_{th} = 3$ dB, $1/\alpha_{se} = 0$ dB, $1/\alpha_{re} = 3.5$ dB, 
$1/\beta_{sd} = 3$ dB, $1/\beta_{sr} = 30$ dB (Case I) or $1/\beta_{rd} = 30$ dB (Case II). 
Dashed straight line represents saturation level when $1/\beta_{sr}\rightarrow \infty$.}
\label{FIG_7}
\vspace{-0.5cm}
\end{figure}

\textbf{Effect of unbalance in the dual-hop links on ESR}:
In Fig. \ref{FIG_7}, ESR of the MRC-SC and SC-MRC scheme is plotted versus average SNR 
for the CSI and NOCSI cases assuming the unbalanced Cases I and II. For Case I, the $x$-axis 
represents $1/\beta_{rd}$ with $1/\beta_{sr}=30$ dB, whereas for 
Case II, it represents $1/\beta_{sr}$ with $1/\beta_{rd}=30$ dB.
It can be observed that CSI helps to improve the ESR; however, CSI is 
more beneficial at low SNR. At high SNR, the benefit of using CSI is marginal or 
no improvement can be obtained when compared to NOCSI. 
It can also be observed that the ESR curves saturate to a constant value in Case II. 
The saturation constant is plotted using a horizontal dashed line  with the help of (\ref{eq_39}). 
In Case I, curves do not saturate to a fixed value as in Case II. 
An increase in $1/\beta_{sr}$ increases the probability of successful decoding; however, 
the SNR at $D$ is still constrained by the $R$-$D$ link quality if $1/\beta_{rd}$ is fixed; 
hence, ESR saturates in Case II.  This is similar to the SOP performance in Fig. \ref{FIG_5}.
If $1/\beta_{sr}$ is fixed to a relatively high value, the signal 
is decoded correctly at $R$ and by increasing $1/\beta_{rd}$, the 
ESR can be increased. Hence, ESR does not saturate in Case I. 
By observing Fig. \ref{FIG_5} and Fig. \ref{FIG_7}, it can be concluded that 
the two unbalanced cases do not yield symmetric results. The figure clearly depicts that ESR can be negative 
if CSI is not available at the transmitters.

\section{Conclusion}  
\label{sec_conclusion}
In this paper, the effects caused by the CSI knowledge on the SOP and ESR have been
studied for a dual-hop cooperative system with a threshold-selection DF relay. 
Combinations of MRC and SC diversity schemes at the destination and eavesdropper were employed 
to take advantage of the direct links. The threshold-selection relay model 
can generalize perfect decoding and wiretap channel results.
Closed-form expressions were derived and the asymptotic 
analysis was presented when the dual-hop link SNRs were balanced and unbalanced. It is found 
that the unbalanced cases become the performance bottleneck; however, 
their effect is not symmetric. It has been observed that knowledge 
of CSI can provide an improved performance; however, 
the degree of improvement is higher at a lower required rate for the SOP and 
at a lower operating SNR for the ESR. It is also concluded that the secrecy is more sensitive 
with the changes in source to eavesdropper link quality.

\section*{Acknowledgment}
\label{sec_acknowledgment}
This work was supported in part by the Royal Society-SERB Newton International Fellowship under Grant NF151345, 
the U.K. Engineering and Physical Sciences Research Council under Grant EP/P019374/1,
the DST-SERB under National Post Doctoral Fellowship Grant PDF/2016/003637, and the 
Natural Science and Engineering Council of Canada (NSERC), through its Discovery program.

% \bibliographystyle{IEEEtran}
% \bibliography{IEEEabrv,MYALL_REFERENCE}

\newcommand{\noop}[1]{}

% \afterpage{\clearpage}

% \begin{comment}
% % TABLE 1
% \setstretch{0.8}
% \begin{longtable} {m{\textwidth}}
\begin{table*}[!t]
\vspace*{-2.5in}
\caption{SOP when CSI is unavailable at the transmitters.} 
\label{table_SOP_outg}
\begin{tabular}{m{\textwidth}}
\\
\hline
\hline
\\
{\centering  SOP of the MRC-SC scheme. 
\begin{align}
\label{eq_19}
P_o(R_s)
&=\lb(1-e^{-\beta_{sr}\gamma_{th}}\rb)
\lb(1-\frac{\alpha_{se}e^{-\beta_{sd}\lb(\rho-1\rb)}}{\alpha_{se}+\rho\beta_{sd}}\rb)  
+e^{-\beta_{sr}\gamma_{th}}\lb[1 \rb. \nn \\ 
&\lb.
+\frac{\beta_{rd}\lb(\alpha_{se}+\alpha_{re}\rb)e^{-\beta_{sd}(\rho-1)}}
{\lb(\beta_{rd}-\beta_{sd}\rb)\lb(\alpha_{se}+\alpha_{re}+\rho\beta_{sd}\rb)} -\frac{\beta_{rd}\lb(\rho\alpha_{se}\beta_{sd}+\rho\alpha_{re}\beta_{sd}+2\alpha_{se}\alpha_{re}\rb)e^{-\beta_{sd}(\rho-1)}}
{\lb(\beta_{rd}-\beta_{sd}\rb)\lb(\alpha_{se}+\rho\beta_{sd}\rb)\lb(\alpha_{re}+\rho\beta_{sd}\rb)} 
\rb. \nn \\ 
&\lb.
-\frac{\beta_{sd}e^{-\beta_{rd}(\rho-1)}}{\lb(\beta_{sd}-\beta_{rd}\rb)}
\lb(\frac{\rho\alpha_{se}\beta_{rd}+\rho\alpha_{re}\beta_{rd}+2\alpha_{se}\alpha_{re}}
{\lb(\alpha_{se}+\rho\beta_{rd}\rb)\lb(\alpha_{re}+\rho\beta_{rd}\rb)} -\frac{\lb(\alpha_{se}+\alpha_{re}\rb)}
{\alpha_{se}+\alpha_{re}+\rho\beta_{rd}}\rb)
\rb].
\end{align}
}
\\
\hline
\\
{
\centering SOP of the MRC-MRC scheme. 
\begin{align}
\label{eq_20}
P_o(R_s)
&=\lb(1-e^{-\beta_{sr}\gamma_{th}}\rb)
\lb(1-\frac{\alpha_{se}e^{-\beta_{sd}\lb(\rho-1\rb)}}{\alpha_{se}+\rho\beta_{sd}}\rb)
+e^{-\beta_{sr}\gamma_{th}}\lb[1 
-\frac{\alpha_{se}\alpha_{re}}{\lb(\beta_{sd}-\beta_{rd}\rb)} \times \rb. \nn \\
&\lb.
\lb(\frac{\beta_{sd}e^{-\beta_{rd}(\rho-1)}}
{\lb(\alpha_{se}+\rho\beta_{rd}\rb)\lb(\alpha_{re}+\rho\beta_{rd}\rb)} -\frac{\beta_{rd}e^{-\beta_{sd}(\rho-1)}}
{\lb(\alpha_{se}+\rho\beta_{sd}\rb)\lb(\alpha_{re}+\rho\beta_{sd}\rb)}\rb)\rb].
\end{align}
}
\\
\hline
\\
{\centering SOP of the SC-SC scheme.  
\begin{align}
\label{eq_21}
P_o(R_s)
&=\lb(1-e^{-\beta_{sr}\gamma_{th}}\rb)
\lb(1 -\frac{\alpha_{se}e^{-\beta_{sd}\lb(\rho-1\rb)}}{\alpha_{se}+\rho\beta_{sd}}\rb)\nn\\
&+e^{-\beta_{sr}\gamma_{th}}
\lb[1-\alpha_{se}\alpha_{re}\lb(\frac{\lb(\alpha_{se}+\alpha_{re}+2\rho\beta_{sd}\rb)e^{-\beta_{sd}\lb(\rho-1\rb)}}
{\lb(\alpha_{se}+\rho\beta_{sd}\rb)\lb(\alpha_{re}+\rho\beta_{sd}\rb)\lb(\alpha_{se}+\alpha_{re}+\rho\beta_{sd}\rb)} \rb.\rb. \nn \\
&\lb.\lb.+\frac{\lb(\alpha_{se}+\alpha_{re}+2\rho\beta_{rd}\rb)e^{-\beta_{rd}(\rho-1)}}
{\lb(\alpha_{se}+\rho\beta_{rd}\rb)\lb(\alpha_{re}+\rho\beta_{rd}\rb)\lb(\alpha_{se}+\alpha_{re}+\rho\beta_{rd}\rb)} \rb.\rb. \nn \\
&\lb.\lb.-\frac{\lb(\alpha_{se}+\alpha_{re}+2\rho\lb(\beta_{sd}+\beta_{rd}\rb)\rb)e^{-\lb(\beta_{sd}+\beta_{rd}\rb)(\rho-1)}}
{\lb(\alpha_{se}+\rho\beta_{sd}+\rho\beta_{rd}\rb)\lb(\alpha_{re}+\rho\beta_{sd}+\rho\beta_{rd}\rb)
\lb(\alpha_{se}+\alpha_{re}+\rho\beta_{sd}+\rho\beta_{rd}\rb)}\rb)
\rb].
\end{align}
}
\\
\hline
\\
{\centering SOP of the SC-MRC scheme.  
\begin{align}
\label{eq_22}
P_o(R_s)
&=\lb(1-e^{-\beta_{sr}\gamma_{th}}\rb)
\lb(1-\frac{\alpha_{se}e^{-\beta_{sd}\lb(\rho-1\rb)}}{\alpha_{se}+\rho\beta_{sd}}\rb)
+e^{-\beta_{sr}\gamma_{th}}
\lb[1-\alpha_{se}\alpha_{re}\lb(\frac{e^{-\beta_{sd}(\rho-1)}}
{\lb(\alpha_{se}+\rho\beta_{sd}\rb)\lb(\alpha_{re}+\rho\beta_{sd}\rb)} \rb.\rb. \nn \\
&\lb.\lb.+\frac{e^{-\beta_{rd}(\rho-1)}} 
{\lb(\alpha_{se}+\rho\beta_{rd}\rb)\lb(\alpha_{re}+\rho\beta_{rd}\rb)} 
-\frac{e^{-\lb(\beta_{sd}+\beta_{rd}\rb)\lb(\rho-1\rb)}}
{\lb(\alpha_{se}+\rho\lb(\beta_{sd}+\beta_{rd}\rb)\rb)\lb(\alpha_{re}+\rho\lb(\beta_{sd}+\beta_{rd}\rb)\rb)}\rb)\rb].
\end{align}
}
\\
\hline
\hline
\\
\end{tabular}
\end{table*}
% \end{longtable}

% % TABLE 2
% \begin{longtable}
% [H]
% {m{\textwidth}}
\begin{table*}[t]
\vspace*{-15mm}
\caption{SOP when CSI is available at the transmitters.} 
\label{table_SOP_csi}
\begin{tabular}{m{\textwidth}}
\\
\hline
\hline
\\
{\centering SOP of the MRC-SC scheme. 
\begin{align}
\label{eq_25}
&P_o(R_s)
=\lb(1-e^{-\beta_{sr}\gamma_{th}}\rb)\lb(\frac{\alpha_{se}}{\alpha_{se}+\beta_{sd}}
-\frac{\alpha_{se}e^{-\beta_{sd}\lb(\rho-1\rb)}}{\alpha_{se}+\rho\beta_{sd}} \rb) 
+e^{-\beta_{sr}\gamma_{th}}\lb[\frac{\beta_{sd}\alpha_{se}\alpha_{re}}{\beta_{sd}-\beta_{rd}} \times \rb.\nn \\
&\lb.\lb(\frac{\alpha_{se}+\alpha_{re}+2\beta_{rd}}
{\lb(\alpha_{se}+\beta_{rd}\rb)\lb(\alpha_{re}+\beta_{rd}\rb)\lb(\alpha_{se}+\alpha_{re}+\beta_{rd}\rb)} 
-\frac{\lb(\alpha_{se}+\alpha_{re}+2\rho\beta_{rd}\rb)e^{-\beta_{rd}(\rho-1)}}
{\lb(\alpha_{se}+\rho\beta_{rd}\rb)\lb(\alpha_{re}+\rho\beta_{rd}\rb)\lb(\alpha_{se}+\alpha_{re}+\rho\beta_{rd}\rb)} \rb) \rb. \nn \\
&\lb.
+\frac{\beta_{rd}\alpha_{se}\alpha_{re}}{\beta_{rd}-\beta_{sd}}
\lb(\frac{\alpha_{se}+\alpha_{re}+2\beta_{sd}}
{\lb(\alpha_{se}+\beta_{sd}\rb)\lb(\alpha_{re}+\beta_{sd}\rb)\lb(\alpha_{se}+\alpha_{re}+\beta_{sd}\rb)} \rb.\rb. \nn \\
 &\lb.\lb.
-\frac{\lb(\alpha_{se}+\alpha_{re}+2\rho\beta_{sd}\rb)e^{-\beta_{sd}(\rho-1)}}
{\lb(\alpha_{se}+\rho\beta_{sd}\rb)\lb(\alpha_{re}+\rho\beta_{sd}\rb)\lb(\alpha_{se}+\alpha_{re}+\rho\beta_{sd}\rb)}
\rb)\rb] .
\end{align}
} 
\\
\hline
\\
{\centering SOP of the MRC-MRC scheme. 
\begin{align}
\label{eq_26}
P_o(R_s)
&=\lb(1-e^{-\beta_{sr}\gamma_{th}}\rb)\lb(\frac{\alpha_{se}}{\alpha_{se}+\beta_{sd}}
-\frac{\alpha_{se}e^{-\beta_{sd}\lb(\rho-1\rb)}}{\alpha_{se}+\rho\beta_{sd}} \rb) \nn \\
&
+e^{-\beta_{sr}\gamma_{th}}
\lb[\frac{\beta_{sd}\alpha_{se}\alpha_{re}}{\lb(\beta_{sd}-\beta_{rd}\rb)\lb(\alpha_{se}
+\beta_{rd}\rb)\lb(\alpha_{re}+\beta_{rd}\rb)} -\frac{\beta_{sd}\alpha_{se}\alpha_{re}e^{-\beta_{rd}(\rho-1)}}
{\lb(\beta_{sd}-\beta_{rd}\rb)\lb(\alpha_{se}+\rho\beta_{rd}\rb)\lb(\alpha_{re}+\rho\beta_{rd}\rb)} \rb. \nn \\
&\lb.
+\frac{\beta_{rd}\alpha_{se}\alpha_{re}}
{\lb(\beta_{rd}-\beta_{sd}\rb)\lb(\alpha_{se}+\beta_{sd}\rb)\lb(\alpha_{re}+\beta_{sd}\rb)}  
-\frac{\beta_{rd}\alpha_{se}\alpha_{re}e^{-\beta_{sd}(\rho-1)}}
{\lb(\beta_{rd}-\beta_{sd}\rb)\lb(\alpha_{se}+\rho\beta_{sd}\rb)\lb(\alpha_{re}+\rho\beta_{sd}\rb)}\rb].
\end{align}
}
\\
\hline
\\
{\centering SOP of the SC-SC scheme.  
\begin{align}
\label{eq_27}
P_o(R_s)
&=\lb(1-e^{-\beta_{sr}\gamma_{th}}\rb)\lb(\frac{\alpha_{se}}{\alpha_{se}+\beta_{sd}}
-\frac{\alpha_{se}e^{-\beta_{sd}\lb(\rho-1\rb)}}{\alpha_{se}+\rho\beta_{sd}} \rb)
+e^{-\beta_{sr}\gamma_{th}}
\lb[\alpha_{se}\alpha_{re} \times \rb . \nn \\
&\lb.
\lb(\frac{\alpha_{se}+\alpha_{re}+2\beta_{sd}}
{\lb(\alpha_{se}+\beta_{sd}\rb)\lb(\alpha_{re}+\beta_{sd}\rb)\lb(\alpha_{se}+\alpha_{re}+\beta_{sd}\rb)}
+\frac{\alpha_{se}+\alpha_{re}+2\beta_{rd}}
{\lb(\alpha_{se}+\beta_{rd}\rb)\lb(\alpha_{re}+\beta_{rd}\rb)\lb(\alpha_{se}+\alpha_{re}+\beta_{rd}\rb)}\rb.\rb. \nn \\
&\lb.\lb.
-\frac{\alpha_{se}+\alpha_{re}+2\lb(\beta_{sd}+\beta_{rd}\rb)}
{\lb(\alpha_{se}+\beta_{sd}+\beta_{rd}\rb)
\lb(\alpha_{re}+\lb(\beta_{sd}+\beta_{rd}\rb)\rb)
\lb(\alpha_{se}+\alpha_{re}+\beta_{sd}+\beta_{rd}\rb)} \rb.\rb. \nn \\
&\lb.\lb.-\frac{\lb(\alpha_{se}+\alpha_{re}+2\rho\beta_{sd}\rb)e^{-\beta_{sd}(\rho-1)}}
{\lb(\alpha_{se}+\rho\beta_{sd}\rb)\lb(\alpha_{re}+\rho\beta_{sd}\rb)
\lb(\alpha_{se}+\alpha_{re}+\rho\beta_{sd}\rb)} \rb.\rb. \nn \\
&\lb.\lb.
-\frac{\lb(\alpha_{se}+\alpha_{re}+2\rho\beta_{rd}\rb)e^{-\beta_{rd}(\rho-1)}}
{\lb(\alpha_{se}+\rho\beta_{rd}\rb)\lb(\alpha_{re}+\rho\beta_{rd}\rb)
\lb(\alpha_{se}+\alpha_{re}+\rho\beta_{rd}\rb)} \rb.\rb. \nn \\
&\lb.\lb.+\frac{\lb(\alpha_{se}+\alpha_{re}+2\rho\lb(\beta_{sd}+\beta_{rd}\rb)\rb)e^{-\lb(\beta_{sd}+\beta_{rd}\rb)(\rho-1)}}
{\lb(\alpha_{se}+\rho\lb(\beta_{sd}+\beta_{rd}\rb)\rb)
\lb(\alpha_{re}+\rho\lb(\beta_{sd}+\beta_{rd}\rb)\rb)
\lb(\alpha_{se}+\alpha_{re}+\rho\lb(\beta_{sd}+\beta_{rd}\rb)\rb)}\rb)\rb].
\end{align}
}
\\
\hline
\\
{\centering SOP of the SC-MRC scheme.  
\begin{align}
\label{eq_28}
P_o(R_s)
&=\lb(1-e^{-\beta_{sr}\gamma_{th}}\rb)\lb(\frac{\alpha_{se}}{\alpha_{se}+\beta_{sd}}
-\frac{\alpha_{se}e^{-\beta_{sd}\lb(\rho-1\rb)}}{\alpha_{se}+\rho\beta_{sd}} \rb)
+e^{-\beta_{sr}\gamma_{th}}
\lb[\lb(\rho-1\rb)\lb(\frac{\alpha_{se}+\alpha_{re}+\alpha_{se}\alpha_{re}}{\alpha_{se}\alpha_{re}}\rb)\rb. \nn \\
&\lb.+\frac{\alpha_{se}\alpha_{re}}{\lb(\alpha_{se}+\beta_{sd}\rb)\lb(\alpha_{re}+\beta_{sd}\rb)}  
\lb(\frac{1}{\alpha_{se}+\beta_{sd}}+\frac{1}{\alpha_{re}+\beta_{sd}}\rb)
-\frac{\alpha_{se}\alpha_{re}e^{-\beta_{sd}(\rho-1)}}{\lb(\alpha_{se}
+\rho\beta_{sd}\rb)\lb(\alpha_{re}+\rho\beta_{rd}\rb)} \times \rb. \nn\\
&\lb.
\lb(\rho-1+\frac{\rho}{\alpha_{se}+\rho\beta_{sd}} 
+\frac{\rho}{\alpha_{re}+\rho\beta_{sd}}\rb)
\rb].
\end{align}
}
\\
\hline
\hline
\\
\end{tabular}
\end{table*}
% \end{longtable}

% \begin{comment}
% % TABLE 3
% \begin{longtable}
% [H]
% {m{\textwidth}}
\begin{table*}[t]
\caption{ESR of various combinations of diversity combining schemes.} 
\label{table_sec_rate}
% \begin{small}
 \begin{tabular}{m{\textwidth}}
\\
\hline
\hline
\\
{
\centering ESR of the MRC-SC scheme when CSI is unavailable at the transmitters. 
\begin{align}
\label{eq_36}
\bar{C}_s
&=\frac{1}{2\ln2}\lb[\lb(1-e^{-\beta_{sr}\gamma_{th}}\rb)
\lb(e^{\alpha_{se}}\operatorname{Ei}\lb(-\alpha_{se}\rb)-e^{\beta_{sd}}\operatorname{Ei}\lb(-\beta_{sd}\rb)\rb)
+e^{-\beta_{sr}\gamma_{th}} \times\rb. \nn \\
&\lb. \lb(\frac{1}{\beta_{sd}-\beta_{rd}}\lb(\beta_{rd}e^{\beta_{sd}}\operatorname{Ei}\lb(-\beta_{sd}\rb)
-\beta_{sd}e^{\beta_{rd}}\operatorname{Ei}\lb(-\beta_{rd}\rb)\rb) 
+e^{\alpha_{se}}\operatorname{Ei}\lb(-\alpha_{se}\rb)+e^{\alpha_{re}}\operatorname{Ei}\lb(-\alpha_{re}\rb) \rb.\rb.\nn \\
&\lb.\lb.
-e^{\alpha_{se}+\alpha_{re}}\operatorname{Ei}\lb(-\alpha_{se}-\alpha_{re}\rb)\rb) 
\rb].
\end{align}
}
\\
\hline
\\
{
\centering ESR of the MRC-MRC scheme when CSI is unavailable at the transmitters. 
\begin{align}
\label{eq_36_1}
\bar{C}_s
&=\frac{1}{2\ln2}\lb[\lb(1-e^{-\beta_{sr}\gamma_{th}}\rb)
\lb(e^{\alpha_{se}}\operatorname{Ei}\lb(-\alpha_{se}\rb)-e^{\beta_{sd}}\operatorname{Ei}\lb(-\beta_{sd}\rb)\rb)
+e^{-\beta_{sr}\gamma_{th}} \times \rb. \nn \\
&\lb. \lb(\frac{1}{\beta_{sd}-\beta_{rd}}
\lb(\beta_{rd}e^{\beta_{sd}}\operatorname{Ei}\lb(-\beta_{sd}\rb) 
-\beta_{sd}e^{\beta_{rd}}\operatorname{Ei}\lb(-\beta_{rd}\rb) \rb) 
 -\frac{1}{\alpha_{se}-\alpha_{re}}\lb(\alpha_{re}e^{\alpha_{se}}\operatorname{Ei}\lb(-\alpha_{se}\rb) \rb.\rb.\rb. \nn \\
&\lb.\lb.\lb.
-\alpha_{se}e^{\alpha_{re}}\operatorname{Ei}\lb(-\alpha_{re}\rb)\rb)\rb)\rb].
\end{align}
}
\\
\hline
\\
{
\centering ESR of the SC-SC scheme when CSI is unavailable at the transmitters. 
\begin{align}
\label{eq_38}
\bar{C}_s&=\frac{1}{2\ln2}\lb[
\lb(1-e^{-\beta_{sr}\gamma_{th}}\rb)
\lb(e^{\alpha_{se}}\operatorname{Ei}\lb(-\alpha_{se}\rb)-e^{\beta_{sd}}\operatorname{Ei}\lb(-\beta_{sd}\rb)\rb)
+e^{-\beta_{sr}\gamma_{th}}
\lb(e^{\alpha_{se}}\operatorname{Ei}\lb(-\alpha_{se}\rb) \rb.\rb. \nn\\
&\lb.\lb.
+e^{\alpha_{re}}\operatorname{Ei}\lb(-\alpha_{re}\rb)
-e^{\beta_{sd}}\operatorname{Ei}\lb(-\beta_{sd}\rb)
-e^{\beta_{rd}}\operatorname{Ei}\lb(-\beta_{rd}\rb)
+e^{\beta_{sd}+\beta_{rd}}\operatorname{Ei}\lb(-\beta_{sd}-\beta_{rd}\rb) \rb.\rb. \nn\\
&\lb.\lb.-e^{\alpha_{se}+\alpha_{re}}\operatorname{Ei}\lb(-\alpha_{se}-\alpha_{re}\rb)
\rb)\rb].
\end{align}
}
\\
\hline
\\
{
\centering ESR of the SC-MRC scheme when CSI is unavailable at the transmitters.
\begin{align}
\label{eq_42}
\bar{C}_s
&=\frac{1}{2\ln2}\lb[\lb(1-e^{-\beta_{sr}\gamma_{th}}\rb)\lb(e^{\alpha_{se}}\operatorname{Ei}\lb(-\alpha_{se}\rb)
-e^{\beta_{sd}}\operatorname{Ei}\lb(-\beta_{sd}\rb)\rb)
+e^{-\beta_{sr}\gamma_{th}} \times \rb. \nn \\
&\lb.\lb(e^{\beta_{sd}+\beta_{rd}}\operatorname{Ei}\lb(-\beta_{sd}-\beta_{rd}\rb) -e^{\beta_{rd}}\operatorname{Ei}\lb(-\beta_{rd}\rb) 
+\frac{1}{\alpha_{se}-\alpha_{re}}\lb(\alpha_{se}e^{\alpha_{re}}\operatorname{Ei}\lb(-\alpha_{re}\rb)\rb. \rb. \rb. \nn \\
&\lb.\lb. \lb.-\alpha_{re}e^{\alpha_{se}}\operatorname{Ei}\lb(-\alpha_{se}\rb)\rb)\rb)\rb].
\end{align}
}
\\
\hline
\\
{\centering ESR of the MRC-SC scheme when CSI is available at the transmitters.
\begin{align}
\label{eq_37}
\bar{C}_s
&=\frac{1}{2\ln2}\lb[\lb(1-e^{-\beta_{sr}\gamma_{th}}\rb)
\lb(e^{\beta_{sd}+\alpha_{se}}\operatorname{Ei}\lb(-\beta_{sd}-\alpha_{se}\rb)
-e^{\beta_{sd}}\operatorname{Ei}\lb(-\beta_{sd}\rb)\rb)
+\frac{e^{-\beta_{sr}\gamma_{th}}}{\beta_{sd}-\beta_{rd}} \times \rb.\nn \\
&\lb.
\lb(\beta_{rd}e^{\beta_{sd}}\operatorname{Ei}\lb(-\beta_{rd}\rb) 
-\beta_{sd}e^{\beta_{rd}}\operatorname{Ei}\lb(-\beta_{rd}\rb)
+\beta_{sd}e^{\beta_{rd}+\alpha_{se}}\operatorname{Ei}\lb(-\beta_{rd}-\alpha_{se}\rb) 
\rb.\rb.\nn \\
&\lb.\lb.
-\beta_{rd}e^{\beta_{sd}+\alpha_{se}}\operatorname{Ei}\lb(-\beta_{sd}-\alpha_{se}\rb) 
% \rb.\rb.\nn \\
% &\lb.\lb.
+\beta_{sd}e^{\beta_{rd}+\alpha_{re}}\operatorname{Ei}\lb(-\beta_{rd}-\alpha_{re}\rb) 
-\beta_{rd}e^{\beta_{sd}+\alpha_{re}}\operatorname{Ei}\lb(-\beta_{sd}-\alpha_{re}\rb)  \rb.\rb.\nn \\
&\lb.\lb.
+\beta_{rd}e^{\beta_{sd}+\alpha_{se}+\alpha_{re}}\operatorname{Ei}\lb(-\beta_{sd}-\alpha_{se}-\alpha_{re}\rb) 
% \rb.\rb.\nn \\
% &\lb.\lb.
-\beta_{sd}e^{\beta_{rd}+\alpha_{se}+\alpha_{re}}\operatorname{Ei}\lb(-\beta_{rd}-\alpha_{se}-\alpha_{re}\rb) \rb) 
\rb].
\end{align}
}
\\
\hline
\\
{\centering ESR of the MRC-MRC scheme when CSI is available at the transmitters.
\begin{align}
\label{eq_37_1}
\bar{C}_s
&=\frac{1}{2\ln2}\lb[\lb(1-e^{-\beta_{sr}\gamma_{th}}\rb)
\lb(e^{\beta_{sd}+\beta_{rd}}\operatorname{Ei}\lb(-\beta_{sd}-\beta_{rd}\rb)
-e^{\beta_{sd}}\operatorname{Ei}\lb(-\beta_{sd}\rb)\rb)
+\frac{e^{-\beta_{sr}\gamma_{th}}}{\beta_{sd}-\beta_{rd}} \times \rb. \nn \\
&\lb.\lb(\beta_{rd}e^{\beta_{sd}}\operatorname{Ei}\lb(-\beta_{sd}\rb)
% \rb.\rb. \nn \\
% &\lb.\lb.
-\beta_{sd}e^{\beta_{rd}}\operatorname{Ei}\lb(-\beta_{rd}\rb) 
+\frac{1}{\lb(\alpha_{se}-\alpha_{re}\rb)}
\lb(\beta_{sd}\alpha_{se}e^{\beta_{rd}+\alpha_{re}}\operatorname{Ei}\lb(-\beta_{rd}-\alpha_{re}\rb) \rb.\rb.\rb. \nn \\
&\lb.\lb.\lb.
-\beta_{rd}\alpha_{se}e^{\beta_{sd}+\alpha_{re}}\operatorname{Ei}\lb(-\beta_{sd}-\alpha_{re}\rb)
+\beta_{rd}\alpha_{re}e^{\beta_{sd}+\alpha_{se}}\operatorname{Ei}\lb(-\beta_{sd}-\alpha_{se}\rb)
\rb.\rb.\rb. \nn \\
&\lb.\lb.\lb.
-\beta_{sd}\alpha_{re}e^{\beta_{rd}+\alpha_{se}}\operatorname{Ei}\lb(-\beta_{rd}-\alpha_{se}\rb)\rb) \rb)\rb].
\end{align}
}
\\
\hline
\\
\end{tabular}
\end{table*}
% \end{longtable}

\begin{table*}[t]
\vspace*{-106mm}
\begin{tabular}[t]{m{\textwidth}}
\\
\hline
\\
{
% \vspace{-25cm}
\centering ESR of the SC-SC scheme when CSI is available at the transmitters. 
\begin{align}
\label{eq_38_1}
\bar{C}_s&
=\frac{1}{2\ln2}\lb[\lb(1-e^{-\beta_{sr}\gamma_{th}}\rb)
\lb(e^{\beta_{sd}+\beta_{rd}}\operatorname{Ei}\lb(-\beta_{sd}-\beta_{rd}\rb)
-e^{\beta_{sd}}\operatorname{Ei}\lb(-\beta_{sd}\rb)\rb)
+e^{-\beta_{sr}\gamma_{th}} \times \rb.\nn \\
&\lb. \lb(e^{\beta_{sd}+\alpha_{se}}\operatorname{Ei}\lb(-\beta_{sd}-\alpha_{se}\rb)
-e^{\beta_{sd}}\operatorname{Ei}\lb(-\beta_{sd}\rb)
-e^{\beta_{rd}}\operatorname{Ei}\lb(-\beta_{rd}\rb) 
+e^{\beta_{sd}+\beta_{rd}}\operatorname{Ei}\lb(-\beta_{sd}-\beta_{rd}\rb) \rb.\rb. \nn\\
&\lb.\lb.
+e^{\beta_{sd}+\alpha_{re}}\operatorname{Ei}\lb(-\beta_{sd}-\alpha_{re}\rb)  
+e^{\beta_{rd}+\alpha_{se}}\operatorname{Ei}\lb(-\beta_{rd}-\alpha_{se}\rb) 
+e^{\beta_{rd}+\alpha_{re}}\operatorname{Ei}\lb(-\beta_{rd}-\alpha_{re}\rb) \rb.\rb. \nn\\
&\lb.\lb.  
-e^{\beta_{sd}+\alpha_{se}+\alpha_{re}}\operatorname{Ei}\lb(-\beta_{sd}-\alpha_{se}-\alpha_{re}\rb) 
-e^{\beta_{rd}+\alpha_{se}+\alpha_{re}}\operatorname{Ei}\lb(-\beta_{rd}-\alpha_{se}-\alpha_{re}\rb) 
\rb.\rb. \nn\\
&\lb.\lb.
-e^{\beta_{sd}+\beta_{rd}+\alpha_{se}}\operatorname{Ei}\lb(-\beta_{sd}-\beta_{rd}-\alpha_{se}\rb)        
-e^{\beta_{sd}+\beta_{rd}+\alpha_{re}}\operatorname{Ei}\lb(-\beta_{sd}-\beta_{rd}-\alpha_{re}\rb) \rb.\rb. \nn\\
&\lb.\lb.    
+e^{\beta_{sd}+\beta_{rd}+\alpha_{se}+\alpha_{re}} 
\operatorname{Ei}\lb(-\beta_{sd}-\beta_{rd}-\alpha_{se}-\alpha_{re}\rb)        
\rb)\rb].
\end{align}
% \vspace{-10cm}
}
\\
\hline
\\
{
\centering ESR of the SC-MRC scheme when CSI is available at the transmitters.
\begin{align}
\label{eq_43}
\bar{C}_s
&=\frac{1}{2\ln2}\lb[\lb(1-e^{-\beta_{sr}\gamma_{th}}\rb)
\lb(e^{\beta_{sd}+\beta_{rd}}\operatorname{Ei}\lb(-\beta_{sd}-\beta_{rd}\rb)
-e^{\beta_{sd}}\operatorname{Ei}\lb(-\beta_{sd}\rb)\rb)
+e^{-\beta_{sr}\gamma_{th}} \times \rb.\nn \\
&\lb.\lb(\frac{\alpha_{se}}{\alpha_{se}-\alpha_{re}}
\lb(e^{\beta_{sd}+\alpha_{re}}\operatorname{Ei}\lb(-\beta_{sd}-\alpha_{re}\rb)
-e^{\beta_{sd}+\alpha_{se}}\operatorname{Ei}\lb(-\beta_{sd}-\alpha_{se}\rb)
+e^{\beta_{rd}+\alpha_{re}}\operatorname{Ei}\lb(-\beta_{rd}-\alpha_{re}\rb)
\rb)
\rb.\rb.\nn \\
&\lb.\lb. -\frac{\alpha_{re}}{\alpha_{se}-\alpha_{re}}
\lb(e^{\beta_{rd}+\alpha_{se}}\operatorname{Ei}\lb(-\beta_{rd}-\alpha_{se}\rb)
-e^{\beta_{sd}+\beta_{rd}+\alpha_{re}}\operatorname{Ei}\lb(-\beta_{sd}-\beta_{rd}-\alpha_{re}\rb)\rb)
-e^{\beta_{sd}}\operatorname{Ei}\lb(-\beta_{sd}\rb)\rb.\rb.\nn \\
&\lb.\lb.
-e^{\beta_{rd}}\operatorname{Ei}\lb(-\beta_{rd}\rb) +e^{\beta_{sd}+\beta_{rd}}\operatorname{Ei}\lb(-\beta_{sd}-\beta_{rd}\rb)
\rb)\rb]. 
\end{align}
}
\\
\hline
\hline
\\
\end{tabular}
\end{table*}

% % TABLE 4
% \begin{longtable}
% [H]
% {m{\textwidth}}
\begin{table*}[t]
\caption{Asymptotic SOP under the Balanced Case.} 
\label{talbe_SOP_bal}
\begin{tabular}{m{\textwidth}}
\\
\hline
\hline
\\
{\centering Asymptotic SOP of the MRC-SC scheme when CSI is unavailable, derived from (\ref{eq_19}). 
\begin{align}
\label{eq_44}
P_o^{AS}(R_s)
&=\frac{1}{\frac{1}{\beta}}\lb[
\gamma_{th}-1-\frac{1}{\beta_{sd}} 
+\frac{\rho\lb(\alpha_{se}^2+\alpha_{re}^2+\alpha_{se}\alpha_{re}\lb(\alpha_{se}+\alpha_{re}+1\rb)\rb)}
{\alpha_{se}\alpha_{re}\lb(\alpha_{se}+\alpha_{re}\rb)}\rb.\nn \\
&\lb.+\frac{\alpha_{se}e^{-\beta_{sd}(\rho-1)}}{\beta_{sd}\lb(\alpha_{se}+\rho\beta_{sd}\rb)}
\lb(\alpha_{re}\lb(\alpha_{se}+\alpha_{re}+2\rho\beta_{sd}\rb) 
% \rb.\rb.\nn \\
% &\lb.\lb.
-\gamma_{th}\beta_{sd}\rb)\rb].
\end{align}
}
\\
\hline
\\
{\centering Asymptotic SOP of the MRC-MRC scheme when CSI is unavailable, derived from (\ref{eq_20}).
\begin{align}
\label{eq_48}
P_o^{AS}(R_s)
&=\frac{1}{\frac{1}{\beta}}\lb[\gamma_{th}-1-\frac{1}{\beta_{sd}}
+\frac{\rho\lb(\alpha_{se}+\alpha_{re}+\alpha_{se}\alpha_{re}\rb)}{\alpha_{se}\alpha_{re}} \rb.\nn \\
 &\lb.
-\frac{\alpha_{se}e^{-\beta_{sd}\lb(\rho-1\rb)}}{\alpha_{se}+\rho\beta_{sd}} 
\lb(\gamma_{th}-\frac{\alpha_{re}}{\beta_{sd}\lb(\alpha_{re}+\rho\beta_{sd}\rb)}\rb)\rb] .
\end{align}
}
\\
\hline
\\
{\centering Asymptotic SOP of the SC-MRC scheme when CSI is unavailable, derived from (\ref{eq_22}).
\begin{align}
\label{eq_50}
P_o^{AS}(R_s)&=\frac{1}{\frac{1}{\beta}}\lb[\gamma_{th}-1
+\frac{\rho\lb(\alpha_{se}+\alpha_{re}+\alpha_{se}\alpha_{re}\rb)}{\alpha_{se}\alpha_{re}}
-\frac{\alpha_{se}e^{-\beta_{sd}(\rho-1)}}{\alpha_{se}+\rho\beta_{sd}}
\lb(\gamma_{th}+\frac{\alpha_{re}}{\alpha_{re}+\rho\beta_{sd}}\times \rb.\rb. \nn\\
&\lb.\lb.\lb( \rho-1+\frac{\rho}{\alpha_{se}+\rho\beta_{sd}}+
\frac{\rho}{\alpha_{re}+\rho\beta_{sd}}\rb)\rb)
\rb].
\end{align}
}
\\
\hline
\\
{
\centering Asymptotic SOP of the MRC-SC scheme when CSI is available, derived from (\ref{eq_25}). 
\begin{align}
\label{eq_47}
P_o^{AS}(R_s)
&=\frac{1}{\frac{1}{\beta}}\lb[\lb(\rho-1\rb)\lb(1+\frac{1}{\alpha_{se}}+\frac{1}{\alpha_{re}}-\frac{1}{\alpha_{se}+\alpha_{re}}\rb)
+\alpha_{se}\gamma_{th}\lb(\frac{1}{\alpha_{se}+\beta_{sd}}-\frac{e^{-\beta_{sd}(\rho-1)}}{\alpha_{se}+\rho\beta_{sd}}\rb)\rb. \nn \\
&\lb.-\frac{\alpha_{se}\alpha_{re}}{\beta_{sd}}\lb(\frac{\alpha_{se}+\alpha_{re}+2\beta_{sd}}
{(\alpha_{se}+\beta_{sd})(\alpha_{re}+\beta_{sd})(\alpha_{se}+\alpha_{re}+\beta_{sd})} \rb.\rb.\nn \\
 &\lb.\lb.
-\frac{(\alpha_{se}+\alpha_{re}+2\rho\beta_{sd})e^{-\beta_{sd}(\rho-1)}}
{(\alpha_{se}+\rho\beta_{sd})(\alpha_{re}+\rho\beta_{sd})(\alpha_{se}+\alpha_{re}+\rho\beta_{sd})}\rb)
\rb].
\end{align}
}
\\
\hline
\\
{
\centering Asymptotic SOP of the MRC-MRC scheme when CSI is available, derived from (\ref{eq_26}). 
\begin{align}
\label{eq_47_1}
P_o^{AS}(R_s)
&=\frac{1}{\frac{1}{\beta}}\lb[\lb(\rho-1\rb)\lb(1+\frac{1}{\alpha_{se}}+\frac{1}{\alpha_{re}}\rb)
+\alpha_{se}\gamma_{th}\lb(\frac{1}{\alpha_{se}+\beta_{sd}}-\frac{e^{-\beta_{sd}(\rho-1)}}{\alpha_{se}+\rho\beta_{sd}}\rb)\rb. \nn \\
&\lb.-\frac{\alpha_{se}\alpha_{re}}{\beta_{sd}}\lb(\frac{1}
{\lb(\alpha_{se}+\beta_{sd}\rb)\lb(\alpha_{re}+\beta_{sd}\rb)} 
-\frac{e^{-\beta_{sd}(\rho-1)}}{\lb(\alpha_{se}+\rho\beta_{sd}\rb)\lb(\alpha_{re}+\rho\beta_{sd}\rb)}\rb)
\rb].
\end{align}
}
\\
\hline
\\
{\centering Asymptotic SOP of the SC-MRC scheme when CSI is available, derived from (\ref{eq_28}). 
\begin{align}
\label{eq_50_1}
P_o^{AS}(R_s)&=\frac{1}{\frac{1}{\beta}}\lb[\lb(\rho-1\rb)
\lb(1+\frac{1}{\alpha_{se}}+\frac{1}{\alpha_{re}}\rb)
+\frac{\alpha_{se}\alpha_{re}}{\lb(\alpha_{se}+\beta_{sd}\rb)\lb(\alpha_{re}+\beta_{sd}\rb)}
\lb(\frac{1}{\alpha_{se}+\beta_{sd}}+\frac{1}{\alpha_{re}+\beta_{sd}}\rb)\rb. \nn\\
&\lb.-\frac{\alpha_{se}\alpha_{re}e^{-\beta_{sd}(\rho-1)}}
{\lb(\alpha_{se}+\rho\beta_{sd}\rb)\lb(\alpha_{re}+\rho\beta_{sd}\rb)}
\lb(\rho-1+\frac{\rho}{\alpha_{se}+\rho\beta_{sd}}+\frac{\rho}{\alpha_{re}+\rho\beta_{sd}}\rb) \rb.\nn \\
&\lb.
+\alpha_{se}\gamma_{th}
\lb(\frac{1}{\alpha_{se}+\beta_{sd}}-\frac{e^{-\beta_{sd}(\rho-1)}}{\alpha_{se}+\rho\beta_{sd}}\rb)
\rb].
\end{align}
} 
\\
\hline
\hline
\\
\end{tabular}
\end{table*}
% \end{longtable}
% \vspace{2cm}

% % TABLE 5
% \begin{longtable}
% [H]
% {m{\textwidth}}
\begin{table*}[t]
\caption{Asymptotic SOP under the unbalanced Case I.} 
\label{table_SOP_unbal_I}
\begin{tabular}{m{\textwidth}}
\\
\hline
\hline
\\
{\centering Asymptotic SOP of the MRC-SC scheme when CSI is unavailable, derived from (\ref{eq_19}). 
\begin{align}
\label{eq_51}
P_o^{AS}(R_s)
&=\lb(1-e^{-\beta_{sr}\gamma_{th}}\rb)
\lb(1-\frac{\alpha_{se}e^{-\beta_{sd}\lb(\rho-1\rb)}}{\alpha_{se}+\rho\beta_{sd}}\rb)
+\frac{e^{-\beta_{sr}\gamma_{th}}}{\frac{1}{\beta}}\times \nn \\
&
\lb[\frac{\rho\lb(\alpha_{se}^2+\alpha_{re}^2+\alpha_{se}\alpha_{re}\lb(\alpha_{se}+\alpha_{re}+1\rb)\rb)}
{\alpha_{se}\alpha_{re}\lb(\alpha_{se}+\alpha_{re}\rb)}
-1-\frac{1}{\beta_{sd}} 
\rb. \nn \\
&\lb.
+\frac{\alpha_{se}\alpha_{re}\lb(\alpha_{se}+\alpha_{re}+2\rho\beta_{sd}\rb)e^{-\beta_{sd}(\rho-1)}}
{\beta_{sd}\lb(\alpha_{se}+\rho\beta_{sd}\rb)\lb(\alpha_{re}+\rho\beta_{sd}\rb)\lb(\alpha_{se}+\alpha_{re}+\rho\beta_{sd}\rb)}
 \rb].
\end{align}
}
\\
\hline
\\
{\centering Asymptotic SOP of the MRC-MRC scheme when CSI is unavailable, derived from (\ref{eq_20}). 
\begin{align}
\label{eq_51_X}
P_o^{AS}(R_s)
&=\lb(1-e^{-\beta_{sr}\gamma_{th}}\rb)\lb(1-\frac{\alpha_{se}e^{-\beta_{sd}\lb(\rho-1\rb)}}{\alpha_{se}+\rho\beta_{sd}}\rb)
+\frac{e^{-\beta_{sr}\gamma_{th}}}{\frac{1}{\beta}}
\lb[\frac{\rho\lb(\alpha_{se}+\alpha_{re}+\alpha_{se}\alpha_{re}\rb)}{\alpha_{se}\alpha_{re}}-1  \rb.\nn\\
&\lb.+\frac{\alpha_{se}\alpha_{re}e^{-\beta_{sd}\lb(\rho-1\rb)}}
{\beta_{sd}\lb(\alpha_{se}+\rho\beta_{sd}\rb)\lb(\alpha_{re}+\rho\beta_{sd}\rb)}\rb].
\end{align}
}
\\
\hline
\\
{\centering Asymptotic SOP of the SC-MRC scheme when CSI is unavailable, derived from (\ref{eq_22}). 
\begin{align}
\label{eq_52}
P_o^{AS}(R_s)&
=\lb(1-e^{-\beta_{sr}\gamma_{th}}\rb)
\lb(1-\frac{\alpha_{se}e^{-\beta_{sd}\lb(\rho-1\rb)}}{\alpha_{se}+\rho\beta_{sd}}\rb)
+\frac{e^{-\beta_{sr}\gamma_{th}}}{\frac{1}{\beta}}
\lb[\frac{\rho\lb(\alpha_{se}+\alpha_{re}+\alpha_{se}\alpha_{re}\rb)}
{\alpha_{se}\alpha_{re}}-1 \rb.\nn \\
&\lb.-\frac{\alpha_{se}\alpha_{re}
\lb(\rho\lb(\alpha_{se}+\alpha_{re}+2\rho\beta_{sd}\rb)
-\lb(\rho-1\rb)\lb(\alpha_{se}+\rho\beta_{sd}\rb)\lb(\alpha_{re}+\rho\beta_{sd}\rb)\rb)e^{-\beta_{sd}\lb(\rho-1\rb)}}
{\lb(\alpha_{se}+\rho\beta_{sd}\rb)^2\lb(\alpha_{re}+\rho\beta_{sd}\rb)^2}\rb].
\end{align}}
\\
\hline
\\
{\centering Asymptotic SOP of the MRC-SC scheme when CSI is available, derived from (\ref{eq_25}). 
\begin{align}
\label{eq_51_1}
&P_o^{AS}(R_s)=\lb(1-e^{-\beta_{sr}\gamma_{th}}\rb)\lb(\frac{\alpha_{se}}{\alpha_{se}+\beta_{sd}}
-\frac{\alpha_{se}e^{-\beta_{sd}(\rho-1)}}{\alpha_{se}+\rho\beta_{sd}}\rb)
+\frac{e^{-\beta_{sr}\gamma_{th}}}{\frac{1}{\beta}} \times \nn \\
&
\lb[\lb(\rho-1\rb)\lb(1+\frac{1}{\alpha_{se}}+\frac{1}{\alpha_{re}}-\frac{1}{\alpha_{se}+\alpha_{re}}\rb) 
-\frac{\alpha_{se}\alpha_{re}}{\beta_{sd}}\lb(\frac{\alpha_{se}+\alpha_{re}+2\beta_{sd}}
{\lb(\alpha_{se}+\beta_{sd}\rb)\lb(\alpha_{re}+\beta_{sd}\rb)\lb(\alpha_{se}+\alpha_{re}+\beta_{sd}\rb)}
\rb.\rb. \nn \\
&\lb.\lb.
-\frac{\lb(\alpha_{se}+\alpha_{re}+2\rho\beta_{sd}\rb)e^{-\beta_{sd}(\rho-1)}}
{\lb(\alpha_{se}+\rho\beta_{sd}\rb)\lb(\alpha_{re}+\rho\beta_{sd}\rb)\lb(\alpha_{se}+\alpha_{re}+\rho\beta_{sd}\rb)}\rb)
\rb].
\end{align}
}
\\
\hline
\\
{\centering Asymptotic SOP of the MRC-MRC scheme when CSI is available, derived from (\ref{eq_26}). 
\begin{align}
\label{eq_51_Y}
P_o^{AS}(R_s)
&=\lb(1-e^{-\beta_{sr}\gamma_{th}}\rb)\lb(\frac{\alpha_{se}}{\alpha_{se}+\beta_{sd}}
-\frac{\alpha_{se}e^{-\beta_{sd}(\rho-1)}}{\alpha_{se}+\rho\beta_{sd}}\rb)
+\frac{e^{-\beta_{sr}\gamma_{th}}}{\frac{1}{\beta}}
\lb[\lb(\rho-1\rb)\lb(1+\frac{1}{\alpha_{se}}+\frac{1}{\alpha_{re}}\rb) \rb.\nn \\
&\lb.-\frac{\alpha_{se}\alpha_{re}}{\beta_{sd}\lb(\alpha_{se}+\beta_{sd}\rb)\lb(\alpha_{re}+\beta_{sd}\rb)} 
+\frac{\alpha_{se}\alpha_{re}e^{-\beta_{sd}\lb(\rho-1\rb)}}
{\beta_{sd}\lb(\alpha_{se}+\rho\beta_{sd}\rb)\lb(\alpha_{re}+\rho\beta_{sd}\rb)} 
\rb].
\end{align}
}
\\
\hline
\\
{\centering Asymptotic SOP of the SC-MRC scheme when CSI is available, derived from (\ref{eq_28}). 
\begin{align}
\label{eq_52_1}
P_o^{AS}(R_s)&=
\lb(1-e^{-\beta_{sr}\gamma_{th}}\rb)
\lb(\frac{\alpha_{se}}{\alpha_{se}+\beta_{sd}}
-\frac{\alpha_{se}e^{-\beta_{sd}(\rho-1)}}{\alpha_{se}+\rho\beta_{sd}}\rb)
+\frac{e^{-\beta_{sr}\gamma_{th}}}{\frac{1}{\beta}}
\lb[\lb(\rho-1\rb)\lb(1+\frac{1}{\alpha_{se}}+\frac{1}{\alpha_{re}}\rb) \rb. \nn \\
&\lb.
+\frac{\alpha_{se}\alpha_{re}}{\lb(\alpha_{se}+\beta_{sd}\rb)\lb(\alpha_{re}+\beta_{sd}\rb)}
% \times \rb. \nn \\
% &\lb.
\lb(\frac{1}{\alpha_{re}+\beta_{sd}}+\frac{1}{\alpha_{se}+\beta_{sd}}\rb) \rb. \nn \\
&\lb.
-\frac{\alpha_{se}\alpha_{re}e^{-\beta_{sd}(\rho-1)}}
{\lb(\alpha_{se}+\rho\beta_{sd}\rb)\lb(\alpha_{re}+\rho\beta_{sd}\rb)}
\lb(\rho-1+\frac{\rho}{\alpha_{se}+\rho\beta_{sd}}+\frac{\rho}{\alpha_{re}+\rho\beta_{sd}}\rb)
\rb].
\end{align}
}
\\
\hline
\hline
\\
% % --------------------------------------------------------------
\end{tabular}
\end{table*}
% \end{longtable}

% \begin{comment}
% % TABLE 6
% \begin{longtable}
% [H]
% {m{\textwidth}}
\begin{table*} [t]
\vspace*{-2in}
\caption{Asymptotic SOP under the unbalanced Case II.} 
\label{table_SOP_unbal_II}
% \begin{scriptsize}
\begin{tabular}{m{\textwidth}}
% % --------------------------------------------------------------
\\
\hline
\hline
\\
{\centering Asymptotic SOP of the MRC-SC scheme when CSI is unavailable, derived from (\ref{eq_19}).
\begin{align}
\label{eq_54}
P_o^{AS}(R_s)&=
\lb(1-\frac{\beta_{sd}\alpha_{se}e^{-\beta_{rd}(\rho-1)}}
{\lb(\beta_{sd}-\beta_{rd}\rb)\lb(\alpha_{se}+\rho\beta_{rd}\rb)}
-\frac{\beta_{rd}\alpha_{re}e^{-\beta_{sd}(\rho-1)}}
{\lb(\beta_{rd}-\beta_{sd}\rb)\lb(\alpha_{re}+\rho\beta_{sd}\rb)} \rb. \nn \\
&\lb.
+\frac{\beta_{sd}\alpha_{se}e^{-\beta_{rd}(\rho-1)}}
{\lb(\beta_{sd}-\beta_{rd}\rb)\lb(\alpha_{re}+\rho\beta_{sd}\rb)
\lb(\alpha_{se}+\alpha_{re}+\rho\beta_{rd}\rb)} 
\rb.\nn\\
&\lb.
+\frac{\beta_{rd}\alpha_{re}e^{-\beta_{sd}(\rho-1)}}
{\lb(\beta_{rd}-\beta_{sd}\rb)\lb(\alpha_{se}+\rho\beta_{sd}\rb)
\lb(\alpha_{se}+\alpha_{re}+\rho\beta_{sd}\rb)}\rb)
-\frac{\gamma_{th}}{\frac{1}{\beta}}
\lb[\frac{\alpha_{se}e^{-\beta_{sd}\lb(\rho-1\rb)}}{\alpha_{se}+\rho\beta_{sd}} \rb. \nn \\
&\lb.
-\frac{\beta_{sd}\alpha_{se}e^{-\beta_{rd}(\rho-1)}}
{\lb(\beta_{sd}-\beta_{rd}\rb)\lb(\alpha_{se}+\rho\beta_{rd}\rb)} 
% \rb.\nn\\
% &\lb.
-\frac{\beta_{rd}\alpha_{re}e^{-\beta_{sd}(\rho-1)}}
{\lb(\beta_{rd}-\beta_{sd}\rb)\lb(\alpha_{re}+\rho\beta_{sd}\rb)} \rb. \nn \\
&\lb.
+\frac{\beta_{sd}\alpha_{se}e^{-\beta_{rd}(\rho-1)}}
{\lb(\beta_{sd}-\beta_{rd}\rb)\lb(\alpha_{re}+\rho\beta_{sd}\rb)
\lb(\alpha_{se}+\alpha_{re}+\rho\beta_{rd}\rb)} 
\rb.\nn\\
&\lb.
+\frac{\beta_{rd}\alpha_{re}e^{-\beta_{sd}(\rho-1)}}
{\lb(\beta_{rd}-\beta_{sd}\rb)\lb(\alpha_{se}+\rho\beta_{sd}\rb)
\lb(\alpha_{se}+\alpha_{re}+\rho\beta_{sd}\rb)}\rb].  
\end{align}
}
\\
\hline
\\
{\centering Asymptotic SOP of the MRC-MRC scheme when CSI is unavailable, derived from (\ref{eq_20}).
\begin{align}
\label{eq_55_X}
P_o^{AS}(R_s)
&=\lb(1-\frac{\beta_{sd}\alpha_{se}\alpha_{re}e^{-\beta_{rd}\lb(\rho-1\rb)}}
{\lb(\beta_{sd}-\beta_{rd}\rb)\lb(\alpha_{se}+\rho\beta_{rd}\rb)\lb(\alpha_{re}+\rho\beta_{rd}\rb)}
+\frac{\beta_{rd}\alpha_{se}\alpha_{re}e^{-\beta_{sd}\lb(\rho-1\rb)}}
{\lb(\beta_{sd}-\beta_{rd}\rb)\lb(\alpha_{se}+\rho\beta_{sd}\rb)\lb(\alpha_{re}+\rho\beta_{sd}\rb)}\rb)  \nn \\
&
-\frac{\gamma_{th}}{\frac{1}{\beta}}\lb[\frac{\alpha_{se}e^{-\beta_{sd}\lb(\rho-1\rb)}}{\alpha_{se}+\rho\beta_{sd}} 
-\frac{\beta_{sd}\alpha_{se}\alpha_{re}e^{-\beta_{rd}\lb(\rho-1\rb)}}
{\lb(\beta_{sd}-\beta_{rd}\rb)\lb(\alpha_{se}+\rho\beta_{rd}\rb)\lb(\alpha_{re}+\rho\beta_{rd}\rb)} \rb.\nn\\
&\lb.
+\frac{\beta_{rd}\alpha_{se}\alpha_{re}e^{-\beta_{sd}\lb(\rho-1\rb)}}
{\lb(\beta_{sd}-\beta_{rd}\rb)\lb(\alpha_{se}+\rho\beta_{sd}\rb)\lb(\alpha_{re}+\rho\beta_{sd}\rb)}\rb]
\end{align}
}
\\
\hline
\\
{\centering Asymptotic SOP of the SC-MRC scheme when CSI is unavailable, derived from (\ref{eq_22}).
\begin{align}
\label{eq_57}
P_o^{AS}(R_s)&=
1-\alpha_{se}\alpha_{re}\lb(\frac{e^{-\beta_{sd}(\rho-1)}}
{\lb(\alpha_{se}+\rho\beta_{sd}\rb)\lb(\alpha_{re}+\rho\beta_{sd}\rb)} 
+\frac{e^{-\beta_{rd}(\rho-1)}}{\lb(\alpha_{se}+\rho\beta_{rd}\rb)\lb(\alpha_{re}+\rho\beta_{rd}\rb)} \rb. \nn \\
&\lb.
-\frac{e^{-\lb(\beta_{sd}+\beta_{sd}\rb)\lb(\rho-1\rb)}}
{\lb(\alpha_{se}+\rho\beta_{sd}+\rho\beta_{rd}\rb)\lb(\alpha_{re}+\rho\beta_{sd}+\rho\beta_{rd}\rb)}\rb) 
-\frac{\gamma_{th}}{\frac{1}{\beta}}\lb[\frac{\alpha_{se}e^{-\beta_{sd}\lb(\rho-1\rb)}}{\alpha_{se}+\rho\beta_{sd}} \rb. \nn \\
&\lb.
-\alpha_{se}\alpha_{re}\lb(\frac{e^{-\beta_{sd}(\rho-1)}}
{\lb(\alpha_{se}+\rho\beta_{sd}\rb)\lb(\alpha_{re}+\rho\beta_{sd}\rb)} 
+\frac{e^{-\beta_{rd}(\rho-1)}}
{\lb(\alpha_{se}+\rho\beta_{rd}\rb)\lb(\alpha_{re}+\rho\beta_{rd}\rb)}\rb.\rb. \nn \\
&\lb.\lb.
-\frac{e^{-\lb(\beta_{sd}+\beta_{sd}\rb)\lb(\rho-1\rb)}}
{\lb(\alpha_{se}+\rho\beta_{sd}+\rho\beta_{rd}\rb)\lb(\alpha_{re}+\rho\beta_{sd}+\rho\beta_{rd}\rb)}\rb)\rb]
\end{align}
}
\\
\hline
\\
\end{tabular}
% \end{scriptsize}
\end{table*}
% \end{longtable}

\begin{table*}[t]
\begin{tabular}{m{\textwidth}}
\\
\hline
\\
{\centering Asymptotic SOP of the MRC-SC scheme when CSI is available, derived from (\ref{eq_25}).
\begin{align}
\label{eq_54_1}
P_o^{AS}(R_s)&=
\frac{\beta_{sd}\alpha_{se}\alpha_{re}}{\beta_{sd}-\beta_{rd}}
\lb(\frac{\alpha_{se}+\alpha_{re}+2\beta_{rd}}
{\lb(\beta_{rd}+\alpha_{se}\rb)\lb(\beta_{rd}+\alpha_{re}\rb)\lb(\beta_{rd}+\alpha_{se}+\alpha_{re}\rb)} \rb. \nn \\
&\lb.
-\frac{\lb(\alpha_{se}+\alpha_{re}+2\rho\beta_{rd}\rb)e^{-\beta_{rd}(\rho-1)}}
{\lb(\alpha_{se}+\rho\beta_{rd}\rb)\lb(\alpha_{re}+\rho\beta_{rd}\rb)\lb(\alpha_{se}+\alpha_{re}+\rho\beta_{rd}\rb)} \rb) \nn \\
&
-\frac{\beta_{rd}\alpha_{se}\alpha_{re}}{\beta_{rd}-\beta_{sd}}
\lb(\frac{\alpha_{se}+\alpha_{re}+2\beta_{sd}}
{\lb(\beta_{sd}+\alpha_{se}\rb)\lb(\beta_{sd}+\alpha_{re}\rb)\lb(\beta_{sd}+\alpha_{se}+\alpha_{re}\rb)}
\rb. \nn \\ 
&\lb.
-\frac{\lb(\alpha_{se}+\alpha_{re}+2\rho\beta_{sd}\rb)e^{-\beta_{sd}(\rho-1)}}
{\lb(\alpha_{se}+\rho\beta_{sd}\rb)\lb(\alpha_{re}+\rho\beta_{sd}\rb)\lb(\alpha_{se}+\alpha_{re}+\rho\beta_{sd}\rb)}
\rb)  \nn \\
&
-\frac{\gamma_{th}}{\frac{1}{\beta}}
\lb[\frac{\alpha_{se}e^{-\beta_{sd}\lb(\rho-1\rb)}}{\alpha_{se}+\rho\beta_{sd}}-\frac{\alpha_{se}}{\alpha_{se}+\beta_{sd}} \rb. \nn \\
&\lb.
+\frac{\beta_{sd}\alpha_{se}\alpha_{re}}{\beta_{sd}-\beta_{rd}}
\lb(\frac{\alpha_{se}+\alpha_{re}+2\beta_{rd}}
{\lb(\beta_{rd}+\alpha_{se}\rb)\lb(\beta_{rd}+\alpha_{re}\rb)\lb(\beta_{rd}+\alpha_{se}+\alpha_{re}\rb)} 
\rb.\rb. \nn \\
&\lb.\lb.
-\frac{\lb(\alpha_{se}+\alpha_{re}+2\rho\beta_{rd}\rb)e^{-\beta_{rd}(\rho-1)}}
{\lb(\alpha_{se}+\rho\beta_{rd}\rb)\lb(\alpha_{re}+\rho\beta_{rd}\rb)\lb(\alpha_{se}+\alpha_{re}+\rho\beta_{rd}\rb)} \rb)  \rb. \nn \\
&\lb.
-\frac{\beta_{rd}\alpha_{se}\alpha_{re}}{\beta_{rd}-\beta_{sd}}
\lb(\frac{\alpha_{se}+\alpha_{re}+2\beta_{sd}}
{\lb(\beta_{sd}+\alpha_{se}\rb)\lb(\beta_{sd}+\alpha_{re}\rb)\lb(\beta_{sd}+\alpha_{se}+\alpha_{re}\rb)}\rb.\rb. \nn \\ 
&\lb.\lb.-\frac{\lb(\alpha_{se}+\alpha_{re}+2\rho\beta_{sd}\rb)e^{-\beta_{sd}(\rho-1)}}
{\lb(\alpha_{se}+\rho\beta_{sd}\rb)\lb(\alpha_{re}+\rho\beta_{sd}\rb)\lb(\alpha_{se}+\alpha_{re}+\rho\beta_{sd}\rb)}
\rb)
\rb].
\end{align}
}
\\
\hline
\\
{\centering Asymptotic SOP of the MRC-MRC scheme when CSI is available, derived from (\ref{eq_26}).
\begin{align}
\label{eq_56_X}
P_o^{AS}(R_s)&= 
\lb(\frac{\beta_{sd}\alpha_{se}\alpha_{re}}
{\lb(\beta_{sd}-\beta_{rd}\rb)\lb(\alpha_{se}+\beta_{rd}\rb)\lb(\alpha_{re}+\beta_{rd}\rb)}
-\frac{\beta_{rd}\alpha_{se}\alpha_{re}}
{\lb(\beta_{sd}-\beta_{rd}\rb)\lb(\alpha_{se}+\beta_{sd}\rb)\lb(\alpha_{re}+\beta_{sd}\rb)} \rb.\nn \\
&\lb.-\frac{\beta_{sd}\alpha_{se}\alpha_{re}e^{-\beta_{rd}(\rho-1)}}
{\lb(\beta_{sd}-\beta_{rd}\rb)\lb(\alpha_{se}+\rho\beta_{rd}\rb)\lb(\alpha_{re}+\rho\beta_{rd}\rb)}
+\frac{\beta_{rd}\alpha_{se}\alpha_{re}e^{-\beta_{sd}(\rho-1)}}
{\lb(\beta_{sd}-\beta_{rd}\rb)\lb(\alpha_{se}+\rho\beta_{sd}\rb)\lb(\alpha_{re}+\rho\beta_{sd}\rb)}\rb)  \nn \\
&
-\frac{\gamma_{th}}{\frac{1}{\beta}}
\lb[\frac{\alpha_{se}e^{-\beta_{sd}\lb(\rho-1\rb)}}{\alpha_{se}+\rho\beta_{sd}}
-\frac{\alpha_{se}}{\alpha_{se}+\beta_{sd}} 
+\frac{\beta_{sd}\alpha_{se}\alpha_{re}}
{\lb(\beta_{sd}-\beta_{rd}\rb)\lb(\alpha_{se}+\beta_{rd}\rb)\lb(\alpha_{re}+\beta_{rd}\rb)} \rb. \nn\\
&\lb.
-\frac{\beta_{rd}\alpha_{se}\alpha_{re}}
{\lb(\beta_{sd}-\beta_{rd}\rb)\lb(\alpha_{se}+\beta_{sd}\rb)\lb(\alpha_{re}+\beta_{sd}\rb)} 
-\frac{\beta_{sd}\alpha_{se}\alpha_{re}e^{-\beta_{rd}(\rho-1)}}
{\lb(\beta_{sd}-\beta_{rd}\rb)\lb(\alpha_{se}+\rho\beta_{rd}\rb)\lb(\alpha_{re}+\rho\beta_{rd}\rb)} \rb.\nn \\
&\lb.
+\frac{\beta_{rd}\alpha_{se}\alpha_{re}e^{-\beta_{sd}(\rho-1)}}
{\lb(\beta_{sd}-\beta_{rd}\rb)\lb(\alpha_{se}+\rho\beta_{sd}\rb)\lb(\alpha_{re}+\rho\beta_{sd}\rb)}
\rb]
\end{align}
}
\\
\hline
\\
{\centering Asymptotic SOP of the SC-MRC scheme when CSI is available, derived from (\ref{eq_28})
\begin{align}
\label{eq_57_1}
P_o^{AS}(R_s)&=\frac{\alpha_{se}\alpha_{re}e^{-\beta_{sd}(\rho-1)}}
{\lb(\alpha_{se}+\rho\beta_{sd}\rb)\lb(\alpha_{re}+\rho\beta_{sd}\rb)}
-\frac{\alpha_{se}\alpha_{re}}{\lb(\alpha_{se}+\beta_{sd}\rb)\lb(\alpha_{re}+\beta_{sd}\rb)} 
-\frac{\alpha_{se}\alpha_{re}}{\lb(\alpha_{se}+\beta_{rd}\rb)\lb(\alpha_{re}+\beta_{rd}\rb)} \nn \\
&
 +\frac{\alpha_{se}\alpha_{re}e^{-\beta_{rd}(\rho-1)}}{\lb(\alpha_{se}+\rho\beta_{rd}\rb)\lb(\alpha_{re}+\rho\beta_{rd}\rb)} 
 +\frac{\alpha_{se}\alpha_{re}}{\lb(\alpha_{se}+\beta_{sd}+\beta_{rd}\rb)\lb(\alpha_{re}+\beta_{sd}+\beta_{rd}\rb)} \nn \\
&
-\frac{\alpha_{se}\alpha_{re}e^{-\lb(\beta_{sd}+\beta_{sd}\rb)\lb(\rho-1\rb)}}
{\lb(\alpha_{se}+\rho\lb(\beta_{sd}+\beta_{rd}\rb)\rb)\lb(\alpha_{re}+\rho\lb(\beta_{sd}+\beta_{rd}\rb)\rb)} 
-\frac{\gamma_{th}}{\frac{1}{\beta}}
\lb[\frac{\alpha_{se}e^{-\beta_{sd}\lb(\rho-1\rb)}}{\alpha_{se}+\rho\beta_{sd}}
-\frac{\alpha_{se}}{\alpha_{se}+\beta_{sd}} \rb.\nn \\
&\lb.
+\frac{\alpha_{se}\alpha_{re}e^{-\beta_{sd}(\rho-1)}}{\lb(\alpha_{se}+\rho\beta_{sd}\rb)\lb(\alpha_{re}+\rho\beta_{sd}\rb)}
-\frac{\alpha_{se}\alpha_{re}}{\lb(\alpha_{se}+\beta_{sd}\rb)\lb(\alpha_{re}+\beta_{sd}\rb)} 
-\frac{\alpha_{se}\alpha_{re}}{\lb(\alpha_{se}+\beta_{rd}\rb)\lb(\alpha_{re}+\beta_{rd}\rb)} \rb.\nn \\
&\lb.
 +\frac{\alpha_{se}\alpha_{re}e^{-\beta_{rd}(\rho-1)}}{\lb(\alpha_{se}+\rho\beta_{rd}\rb)\lb(\alpha_{re}+\rho\beta_{rd}\rb)} 
 +\frac{\alpha_{se}\alpha_{re}}{\lb(\alpha_{se}+\beta_{sd}+\beta_{rd}\rb)\lb(\alpha_{re}+\beta_{sd}+\beta_{rd}\rb)} \rb.\nn \\
&\lb.
-\frac{\alpha_{se}\alpha_{re}e^{-\lb(\beta_{sd}+\beta_{sd}\rb)\lb(\rho-1\rb)}}
{\lb(\alpha_{se}+\rho\lb(\beta_{sd}+\beta_{rd}\rb)\rb)\lb(\alpha_{re}+\rho\lb(\beta_{sd}+\beta_{rd}\rb)\rb)}
\rb].
\end{align}
}
\\
\hline
\hline
\end{tabular}
% \end{scriptsize}
\end{table*}

\end{document}